\documentclass[12pt, a4paper]{article}
\pdfoutput=1
\usepackage[T1]{fontenc}
\usepackage[utf8]{inputenc}
\usepackage{graphicx}
\usepackage{float}
\usepackage[twoside, inner = 2.5cm, outer = 1.5cm, top = 2.5cm, bottom = 2.5cm]{geometry}
\usepackage{setspace}
\usepackage{amsmath}
\usepackage{amsfonts} 
\usepackage{mathptmx}
\usepackage{xcolor}
\definecolor{bg}{rgb}{0.875, 0.875, 0.875}
\usepackage{multicol}
\renewcommand{\thefootnote}{\fnsymbol{footnote}}
\usepackage{listings}
\usepackage{courier}
\usepackage{titleps}
\newpagestyle{myheader}{
  \sethead{\footnotesize Javier Díaz et al.}{}{\footnotesize Recovering Arrhythmic EEG Transients from Their Stochastic Interference.}
  \setheadrule{.4pt} % Add a rule under the header
  \setfoot{}{\thepage}{}
}
\usepackage{array}
\usepackage{siunitx}
\usepackage{natbib}
\begin{document}
%---------------------------------------------
\begin{center}
{\huge Recovering Arrhythmic EEG Transients from Their Stochastic Interference\footnote[0]{\textbf{Acknowledgements:} We thank Diane Greenstein for editorial assistance. \textbf{Funding}. This work was supported by Japan Society for the Promotion of Science (JSPS) grant 19K12199 (supporting J.D.). \textbf{Author contributions}. J.D. discovered a method for recovering the kernel of a filtered Poisson process and investigated its implications for EEG analysis. H.A. formalized mathematical proofs. J.D. and J-C.L. studied computational models. J.D., G.H., O.M., and K.E.V. devised experimental methods.  G.H., O.M., and X.H. performed surgical procedures. X.H. performed sleep deprivation experiments. J.D. wrote custom code for data analysis, performed data analysis, and prepared figures. J.D. and J-C.L. wrote the manuscript. M.Y. and K.E.V. revised and provided critical feedback on the manuscript. \textbf{Competing interests}. The authors declare that they have no competing interests.}}
\end{center}
\begin{center}
{\normalsize Javier Díaz,$^{1}$\footnote[2]{Corresponding author. Email: diaz.antonio.fn@u.tsukuba.ac.jp; javdiaz@uchile.cl} Hiroyasu Ando,$^{2,3}$ GoEun Han,$^{1}$ Olga Malyshevskaya,$^{1}$ Xifang Hayashi,$^{1}$\\Juan-Carlos Letelier,$^{4}$ Masashi Yanagisawa,$^{1}$ Kaspar E. Vogt$^{1}$}
\end{center}
{
\begin{singlespace}
\setlength{\parindent}{1.00cm}
\indent\indent\indent \scriptsize{$^{1}$International Institute for Integrative Sleep Medicine (IIIS), University of Tsukuba. Tsukuba, Ibaraki 305-8575, Japan.}\\
\indent\indent\scriptsize{$^{2}$Advanced Institute for Materials Research, Tohoku University. Sendai, Miyagi 980-8577, Japan.}\\
\indent\indent\scriptsize{$^{3}$Faculty of Engineering, Information and Systems, Univerisity of Tsukuba. Tsukuba, Ibaraki 305-8573, Japan.}\\
\indent\indent\scriptsize{$^{4}$Departamento de Biología, Facultad de Ciencias, Universidad de Chile. Casilla 653, Santiago 7800003, Chile.}
\end{singlespace}
}
\renewcommand*{\thefootnote}{\arabic{footnote}}

%---------------------------------------------
\setstretch{1.20}
\begin{abstract}
\normalsize
Traditionally, the neuronal dynamics underlying electroencephalograms (EEG) have been understood  as arising from  \textit{rhythmic oscillators with varying degrees of synchronization}. This dominant metaphor employs frequency domain EEG analysis to identify the most prominent populations of neuronal current sources in terms of their frequency and spectral power. However, emerging perspectives on EEG highlight its arrhythmic nature, which is primarily inferred from broadband EEG properties like the ubiquitous $1/f$ spectrum. In the present study, we use an \textit{arrhythmic superposition of pulses} as a metaphor to explain the origin of EEG. This conceptualization has a fundamental problem because the interference produced by the  superpositions of pulses generates colored Gaussian noise, masking the temporal profile of the generating pulse. We solved this problem by developing a mathematical method involving the derivative of the autocovariance function to recover excellent approximations of the underlying pulses, significantly extending the analysis of this type of stochastic processes. When the method is applied to spontaneous mouse EEG sampled at $5$ kHz during the sleep-wake cycle, specific patterns --- called $\Psi$-patterns ---  characterizing NREM sleep, REM sleep, and wakefulness are revealed. $\Psi$-patterns can be understood theoretically as \textit{power density in the time domain} and correspond to combinations of generating pulses at different time scales. Remarkably, we report the first EEG wakefulness-specific feature, which corresponds to an ultra-fast ($\sim 1$ ms) transient component of the observed patterns. By shifting the paradigm of EEG genesis from oscillators to random pulse generators, our theoretical framework pushes the boundaries of traditional Fourier-based EEG analysis, paving the way for new insights into the arrhythmic components of neural dynamics.
\end{abstract}
%---------------------------------------------
\newpage
\pagestyle{myheader}
\setstretch{1.10}
\begin{multicols}{2}
\section*{Introduction}
The neuronal dynamics underlying brain function emerges from complex hierarchical organizational levels. This complexity has traditionally been classified into three levels of organization based on anatomical scales and the experimental paradigms used to access them: micro-scale (single-cell), meso-scale (local neural networks, ensembles), and macro-scale (functional integration of distant brain areas) \citep{buzsaki2016micro, Mastrandrea2017-kk, Hilgetag2020-xt}. Electrophysiology provides a collection of techniques to study neuronal dynamics at these three scales, such as single-cell recordings and patch-clamp techniques, local field potentials (LFPs), and electroencephalography (EEG) \citep{covey2015basic}. EEG, a technique that has been in development for nearly a century, represents macro-scale neuronal dynamics and is a well-established tool in research, clinical practice \citep{Niedermeyer6thEd}, and technological applications, such as brain-machine interface \citep{Rashid2020-at}.

Berger's first human EEG \citep{Berger1929} showed a prominent rhythmic wave activity was observed when the subject was relaxed and closed his or her eyes --- the alpha rhythm ($\sim 10$ Hz) --- while higher frequency components ($> 13$ Hz) were observed to increase in amplitude when the subject opened his or her eyes --- the beta rhythm. Subsequent studies identified some additional EEG rhythms still used today \citep{De_Munck2009-jj}, gamma \citep{Jasper1938}, delta \citep{Walter1936}, etc \citep[chap.~1]{Niedermeyer6thEd}. Since Nobel laureate Edgar Adrian's classic paper \citep{Adrian1934}, these EEG rhythms have been traditionally interpreted as the result of \textit{neuronal oscillator synchronization} at different frequency bands \citep{Buzsaki2012-ev}. Because of this, traditional EEG descriptions emphasize the frequency domain, primarily through the use of Fourier analysis \citep{Campbell2009-lz}. 

The direct connection between Fourier series components and a hypothetical collection of neuronal rhythms has some fundamental flaws. First, a rhythmic conceptualization of EEG disconnects it from what is observed at the micro- and meso-scales, where arrhythmic patterns of transient events (e.g. desynchronized packets of neuronal activity \citep{Luczak2015-yq}) predominate and are frequently found to closely follow Poisson's statistics \citep{Sanger2003-yj}. Second, the vast majority of rhythms experimentally found, across many species and in many neural structures, have asymmetric non-sinusoidal shapes that preclude them from being the orthogonal sinusoidal basis demanded by Fourier analysis \citep{Cole2017-om}. Concerns about a disconnect between neurophysiological processes and their EEG spectral correlates have previously been raised in classic studies \citep{Amzica1998-dl}. Third, the widespread $1/f$ structure found in EEG spectra has been identified as a marker of arrhythmic, scale-free neuronal activity \citep{He2010-zf, Kayser2010-oq, He2014-zk, Donoghue2020-uh, Donoghue2022-hn}. As a result, the fundamental question of whether neural signals (LFPs, EEG) are primarily composed of rhythmic components or transitory events is currently a hot topic of discussion \citep{Van_Ede2018-os, Quinn2019-as, Bauer2020-hn}.

Furthermore, statistical properties of rat EEG during NREM sleep --- fluctuating between phasic activity patterns and Gaussian patterns --- can be reproduced from models interpreting EEG as a stochastic process made by the superposition of neuronal pulsed activity \citep{Diaz2018-mu}. This type of models, technically referred to as \textit{generalized shot noise process} or \textit{Filtered Poisson Process} (FPP), has been used to study a wide range of phenomena such as river overflows \citep{LEFEBVRE20082792}, magnetic instabilities in tokamaks \citep{Garcia2017-cn} and biopotentials \citep{Verveen1974-tn, Fesce1986, Lehky2010-br}. The current study considerably extends the stochastic superposition framework as it presents a novel mathematical result that extracts a good approximation of the process's underlying pulse(s). This approach was used to study spontaneous mouse EEG sampled at $5$ kHz during the sleep-wake cycle revealing novel EEG features resulting from combinations of fast arrhythmic pulses that are specific to different brain states. Our results suggest this approach has wide-reaching implications and applications.
\end{multicols}
%---------------------------------------------
%\newpage
\noindent\rule{\textwidth}{0.4pt}
\begin{multicols}{2}
\section*{Results}
\subsection*{Arrhythmic superposition of pulses.}
To investigate the hypothesis of EEG arising from arrhythmic neuronal activity, let us consider a simple stochastic process $X$ originating from the arrhythmic superposition of $N$ pulses defined by the arbitrary function $g(t)$ (eq. \ref{eq:model}). 

\begin{equation} \label{eq:model}
X(t) = \sum^N_{j=1} g(t - t_j)
\end{equation}

Pulses $g(t)$ carry a finite amount of energy $E_g$ and their time occurrence  ($t_j$) is determined by a homogeneous Poisson process with parameter $\lambda_g$ controlling events per second. It is important to note that $g(t)$ is an energy signal, whereas $X(t)$ is a power signal with power $P_X = \mathit{E_g} \lambda_g$. This simple relationship between the power of $X$ and the energy of the pulse $g(t)$ is based on properties of the Poisson distribution (see supplementary material).

Black traces in fig. \ref{fig:model}A illustrate a sequence of stochastic superpositions $X$ arising from an increasing number of pulses. When the density of pulses $\lambda_g$ is high, the shape of $g(t)$ cannot be recognized within the emerging random patterns. At least apparently, the temporal profile of $g(t)$ seems lost in colored Gaussian noise.  The \textit{spectral color} of $X$ is largely determined by $g(t)$ temporal constant $\tau$, as shown in fig. \ref{fig:model}B (alpha functions with increasing $\tau$ ($2,4,8,16,32$ ms) are properly scaled so they have unit energy fig.\ref{fig:model}B, yellow column, and hence their stochastic superpositions ($X$), at a fixed pulse rate $\lambda_g = 10^4$ pulses per second, have equal power).  Accordingly, the power density plot (fig.\ref{fig:model}B, pink column) shows the spectral redistribution of the same amount of power determined by $\tau$, which is progressively tilted to the lower end of the spectrum as $\tau$ increases. This appears to be a general property because it can be observed for other pulses $g(t)$ with varying temporal widths (supplementary fig. \ref{fig:SpecTilt}).
 
 Figure \ref{fig:model}C illustrates the stationarity of process $X$. Despite the variability among realizations, the time-wise standard deviation of $10^4$ realizations of a process $X$ ($\sigma_X$, red trace) illustrates that the expected amplitude of this Gaussian signal is homogeneous over time.
 
 Given the similarities in the definition of energy and variance of a signal, the variance of $X$ is related to its power as shown in eq. \ref{eq:var&Pow}, where $f_s$ is the sampling rate.

\begin{equation} \label{eq:var&Pow}
\sigma_X^2 = \frac{P_X}{f_s} = \frac{\lambda_g}{f_s} E_g
\end{equation}

%---------------------------------------------------------------------------------
\subsection*{Visualization of statistical properties of process $\boldsymbol{X}$ containing information about the shape of \emph{g(t)}.}
An alternative approach to analyze the power of $X$ is subtracting from $X$ delayed versions of itself referred to as $\hat{X}$ (eq. \ref{eq:Xhat}, note that $t^\prime$ will be used to refer to time delays while $\tau$ will be reserved for exponential decay time constants).

\begin{equation} \label{eq:Xhat}
\hat{X}(t^\prime) = X(t) - X(t + t^\prime)
\end{equation}

\noindent
It can be shown that the variance of $\hat{X}$ is dependent on the delay $t^\prime$ as stated in eq. \ref{eq:var&autocov} (see supplementary material for details).
\begin{equation} \label{eq:var&autocov}
\sigma_{\hat{X}}^2(t^\prime) = 2\frac{\lambda_g}{f_s} \Big( E_g - (g \star g)(t^\prime) \Big)
\end{equation}

\begin{figure}[H]
  \centering
    \includegraphics[width=16.0cm, keepaspectratio]{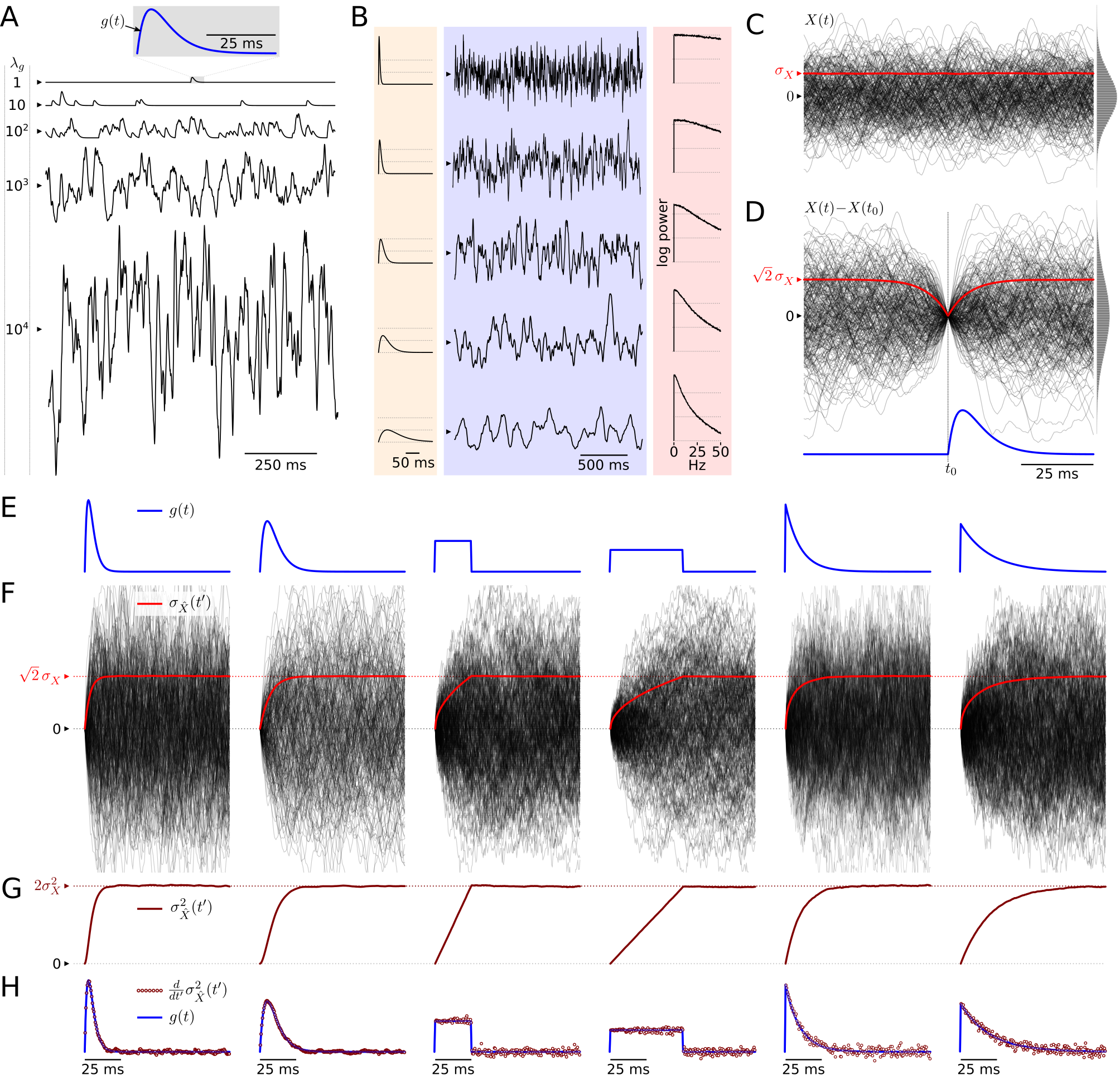}
  \caption{\linespread{1.2} \footnotesize \textbf{Generating a stochastic superposition of the pulse $\boldsymbol{g(t)}$ and then recovering $\boldsymbol{g(t)}$ from the emerging random patterns}.}
  \label{fig:model}
\end{figure}

Due to the \textit{ergodicity}\footnote[1]{A stochastic process is \textit{ergodic} if a sufficiently long realization of the process has the same statistical properties as a large number of realizations of the same process (referred to as an \textit{ensemble}). The standard deviation observed in one realization of $\hat{X}(t^{\prime}) = X(t) - X(t + t^{\prime})$ in our model is equivalent to the standard deviation observed in several realizations of $X_i(t) - X_i(t_0)$ with the same delay $t^{\prime}$ (i.e. time elapsed from $t_0$). Remarkably, the standard deviation between realizations $X_i(t) - X_i(t_0)$ (fig. \ref{fig:model}D red trace) demonstrates a direct relationship between the stochastic process $X(t)$ and its underlying pulse $g(t)$. \citet[p.~47]{Linden2013-qb} provide a rigorous definition of ergodicity.} of process $X$, $\sigma_{\hat{X}}^2(t^\prime)$ can be visualized as in fig. \ref{fig:model}D, where multiple realizations of process $X$ were translated to a common origin at an arbitrary time $t_0$ ($X(t)-X(t_0)$). It is possible to observe a diffusion-like growth takes place from $t_0$, similar to a random walk. The observed amplitude growth can be statistically described using the standard deviation of many realizations of $\hat X$ (fig. \ref{fig:model}D).  Irrespective of statistical variability, $\sigma_{\hat X}(t^\prime)$ can be visualized as growing monotonically, approaching its asymptotic value at $\sqrt{2}\sigma_X$. Importantly, the time it takes for  $\sigma_{\hat X}(t^\prime)$ to reach its asymptotic value matches the time required for $g(t)$ to grow to its maximum and return to zero (blue trace). Notice that while a Dirac delta delivers its energy instantly, the pulse $g(t)$ delivers its energy in a certain finite amount of time (i.e. pulse width). From this perspective, the (variance) of $\hat{X}(t^\prime)$ can be considered a record of the progressive transfer of energy from $g(t)$ to the signal $X$, and thus $\hat{X}(t^\prime)$ becomes stationary when $g(t)$ has delivered all its energy.

Figure \ref{fig:model}F further explores the relationship between $\sigma_{\hat X}(t^\prime)$ and $g(t)$ illustrated by six iso-energetic pulses: two alpha functions with different $\tau$, two square pulses of different length, and two exponentially decaying pulses with different $\tau$. The corresponding $\sigma_{\hat X}(t^\prime)$ are shown in fig. \ref{fig:pulses}G (red traces). Given that these stochastic processes were generated from iso-energetic pulses at the same rate $\lambda_g$, they share the same asymptotic level $\sqrt{2}\sigma_X$, evidencing that despite starting from different pulses the signals have equal power (eq. \ref{eq:var&Pow}). Again (as in fig. \ref{fig:model}D), the temporal profile of $\sigma_{\hat X}$ approaches its asymptotic value when the corresponding pulse has released all its energy. The square pulses, which end abruptly and do not tend asymptotically to zero like the alpha functions, emphasize this idea revealing that at the moment the pulses stop, $\sigma_{\hat X}(t)$ reach its maximum constant level. Figure \ref{fig:model}H shows the profile of the corresponding variance of $\hat X$ ($\sigma^2_{\hat X}(t^\prime)$). Interestingly, $\sigma^2_{\hat X}(t^\prime)$ for the square pulses behaves as perfect ramps, strongly suggesting that $\sigma^2_{\hat X}(t^\prime)$ grows as the integration of $g(t)$. 

%---------------------------------------------------------------------------------
\subsection*{Recovering a pulse from its stochastic interference.}
A remarkable finding of this analysis was revealed when generating $X$ and $\hat{X}$ with classical functions used in neuronal excitation modeling --- e.g. exponential decay, alpha function, and dual-exponential function \citep[chap.~7]{sterratt2011principles}. For these pulses, $\sigma_{\hat{X}}^2(t^\prime)$ also satisfy eq. \ref{eq:fundamentalEq}, where $k$ is the scaling factor defined in eq. \ref{eq:scaling}. Interestingly, this is also true for square pulses.

\begin{equation} \label{eq:fundamentalEq}
\sigma_{\hat{X}}^2(t^\prime) = 2k \int_0^{t^\prime} g(t)dt
\end{equation}

\begin{equation} \label{eq:scaling}
k = \frac{\lambda_g E_g}{f_s \int_0^\infty g(t)dt}
\end{equation}

\noindent
Equation \ref{eq:fundamentalEq}, first conjectured from simulation results, was later mathematically proven for all aforementioned pulse functions (see supplementary material). Importantly, from eq. \ref{eq:fundamentalEq} it follows that the temporal profile of $g(t)$ can be directly obtained differentiating $\sigma_{\hat{X}}^2(t^\prime)$ (eq. \ref{eq:infPulse}), opening the possibility to infer the shape of the underlying pulse $g(t)$ from statistical properties of $X$ (fig. \ref{fig:model}H). Notice that the scale of $g(t)$ cannot be determined from $X$ as the parameters $E_g$ and $\lambda_g$ are inaccessible.
\begin{equation} \label{eq:infPulse}
g(t) \propto \frac{d}{dt^{\prime}} \sigma_{\hat{X}}^2(t^\prime)
\end{equation}

Although important pulse functions related to neuronal dynamics do satisfy eq. \ref{eq:fundamentalEq} (fig. \ref{fig:pulses}A-C), this equation cannot be generalized for arbitrary $g(t)$. However, even for a pulse $g(t)$ not satisfying eq. \ref{eq:fundamentalEq}, the differentiation of $\sigma_{\hat{X}}^2(t^\prime)$ recovers an excellent approximation of $g(t)$ and most importantly, its temporal width (see discussion). Figure \ref{fig:pulses} illustrates examples of recovered pulses belonging to these two conditions: satisfying (upper row) or not satisfying (lower row) eq. \ref{eq:fundamentalEq}. A general relation for $\sigma_{\hat{X}}^2(t^\prime)$ and $g(t)$ is expressed in eq. \ref{eq:generalEq}.
\begin{equation} \label{eq:generalEq}
\sigma_{\hat{X}}^2(t^\prime) = 2\frac{\lambda_g}{f_s} \Big( E_g - (g \star g)(t^\prime) \Big) \simeq 2k \int_0^{t^\prime} g(t)dt
\end{equation}

Interestingly, the inference of $g(t)$ can also be performed from the autocovariance function ($\gamma$) given its relation with $\sigma_{\hat{X}}^2$ according to eq. \ref{eq:autoCov1} (see supplementary material for detail). It is worth mentioning that $\sigma_X^2 = \gamma_X(0)$.
\begin{equation} \label{eq:autoCov1}
\sigma_{\hat{X}}^2(t^\prime) = 2 \Big(\sigma_X^2 - \gamma_X(t^\prime) \Big)
\end{equation}

Combining eq. \ref{eq:autoCov1} with eq. \ref{eq:infPulse} provides two approximations to obtain 
the shape of $g(t)$ from statistical properties of process $X$ (eq. \ref{eq:infPulse2}).
\begin{equation} \label{eq:infPulse2}
g(t) \propto \frac{d}{dt^{\prime}} \sigma_{\hat{X}}^2(t^\prime) \,;\; g(t) \propto -\frac{d}{dt^{\prime}} \gamma_X(t^{\prime})
\end{equation}

Moreover, we define the operator $\Psi$ (eq. \ref{eq:PsiDef}) as the tool to recover a properly scaled approximation of $g(t)$. Conveniently, the scale of $\Psi_X$ is such that its integral corresponds to $\sigma^2_X$ which can be related to the signal power according to eq. \ref{eq:var&Pow}.
\begin{equation} \label{eq:PsiDef}
\Psi_X  = \frac{1}{2} \frac{d}{dt^{\prime}} \sigma_{\hat{X}}^2(t^{\prime}) \,;\; \Psi_X = -\frac{d}{dt^{\prime}} \gamma_X(t^{\prime})
\end{equation}

%---------------------------------------------------------------------------------
\subsection*{$\boldsymbol{\Psi}$ analysis on mixtures of different pulse functions.}
Consider the case of superpositions formed by pulse mixtures with different temporal profiles. In other words process $X$ results from the stochastic superposition, not of one --- as is the case for classic FPP --- but  of two (or more) pulses $g_1(t), g_2(t),...g_n(t)$. Given that the occurrence of pulses of any class is independent, these pulse superpositions can be analyzed by separating each \textit{pulse class} $g_i(t)$. Consequently, the global mixture $X$ can be written as $X = X_1 + X_2 +... + X_n$, where $X_i$ is the homogeneous superposition of all pulses of class $g_i(t)$. When the operator $\Psi$ is used in a mixture $X$ the resulting $\Psi_X$ profile is simply the sum of the individual $[\Psi_{X}]_i$ profiles (eq. \ref{eq:Psi_prop}, see the supplementary material for more information).
\begin{equation} \label{eq:Psi_prop}
\Psi_X = [\Psi_{X}]_1 + [\Psi_{X}]_2 + ... + [\Psi_{X}]_n
\end{equation}

\begin{figure}[H]
  \centering
    \includegraphics[width=12.0cm, keepaspectratio]{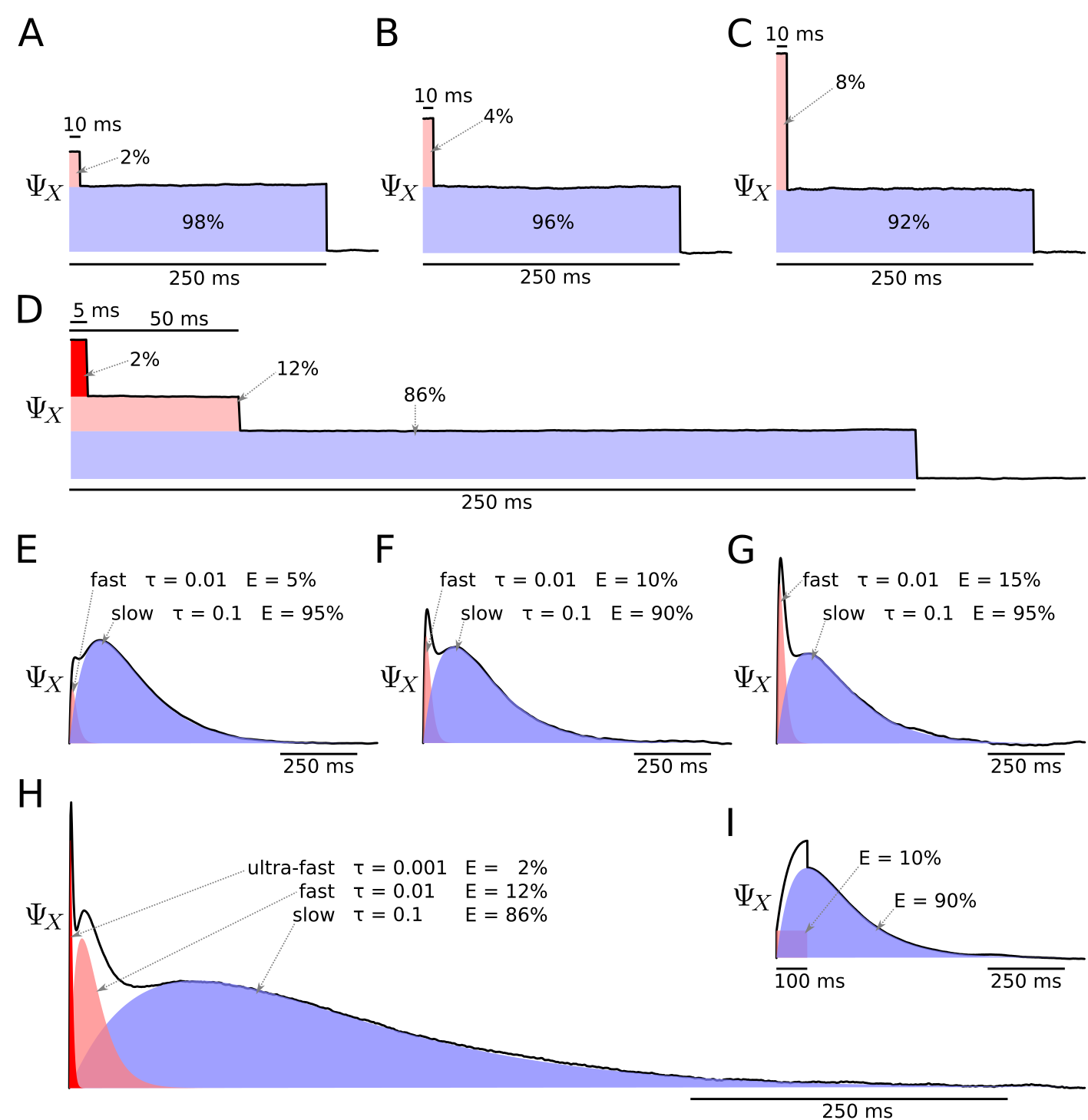}
  \caption{\linespread{1.2} \footnotesize\textbf{$\boldsymbol{\Psi}$ Analysis for the stochastic superposition of different pulses.} Panels \textbf{A-C} show $\Psi$-patterns (black traces) obtained by the analysis of 1-hour signals ($X$) generated combining long square pulses ($250$ ms) with short square pulses ($10$ ms) in a mixture of Poisson processes. The energy contribution of the pulses to the total power is \textbf{A} short, $2\%$; long, $98\%$. \textbf{B} short, $4\%$; long, $96\%$. \textbf{C} short, $8\%$; long, $92\%$. In all three cases, the area under $\Psi$-patterns is naturally divided into two sub-areas (pink and light-blue) which have the same width as the contributing pulses and are in the same proportion as the energy contribution of the corresponding pulses. Panel \textbf{D} shows $\Psi$-pattern for a mixture of 3 square pulses, $5$ ms (red), $50$ ms (pink), and $250$ ms (light-blue) in length, with respective energy contributions: $2\%$, $12\%$, and $86\%$. Panels \textbf{E-G} show $\Psi$-patterns (black traces) obtained from 1-hour signals ($X$) generated combining fast (pink; $\tau = 0.01$ ms) and slow (light-blue; $\tau = 0.1$ ms) alpha functions. The energy (E) contribution of the pulses to the total power is \textbf{E} fast, $5\%$; slow, $95\%$. \textbf{F} fast, $10\%$; slow, $90\%$. \textbf{G} fast, $15\%$; slow, $85\%$. In all three cases, the shape of $\Psi$-patterns is the linear superposition of a fast pulse (pink) and a fast pulse (light-blue) with areas corresponding to their contributions to the total energy of $X$. Panel \textbf{H} shows the analysis of a pulse mixture similar to the one depicted in \textbf{E-G}, but introducing a third faster alpha function, with energy contributions: ultra-fast (red; $\tau = 0.001$ ms), $2\%$; fast (pink; $\tau = 0.01$ ms), $12\%$; slow (light-blue; $\tau = 0.1$ ms), $86\%$. Notice that even small contributions of energy by fast processes can be detected as prominent narrow peaks. Panel \textbf{I} exemplifies the $\Psi$-pattern of a mixture of two pulses of different classes, an alpha function ($\tau = 250$ ms, contributing $90\%$ to the signal power) and a square pulse (width = $100$ ms, contributing $10\%$ to the signal power).}
  \label{fig:pulseMix}
\end{figure}

As was previously discussed, each $[\Psi_{X}]_i$ profile associated with a homogeneous pulse superposition corresponds to an approximation of the underlying pulse $g_i(t)$. Thus, $\Psi_X$ represents the linear superposition of all pulses $g_i(t)$ at play, aligned at $t^\prime = 0$, and $\Psi_X$ total area corresponds to the integrated power contributed from all different $g_i(t)$ pulses (fig. \ref{fig:pulseMix}). Equation \ref{eq:autoCov_sup} shows the relationship between the different pulses $g_i(t)$ and $\Psi_X$, where scaling factors $w_i$ control the area of each pulse reflecting their relative energy contributions:
\begin{equation} \label{eq:autoCov_sup}
\Psi_X \simeq \sum_{i}^N w_i \, g_i(t).
\end{equation}

Empirical results show, for example, the simplicity of $\Psi_X$ when $X$ is a mixture of square pulses of different durations. $\Psi_X$ simply shows the pulses mounted one on top of the other, revealing their temporal duration (fig. \ref{fig:pulseMix}A-D). The area of each rectangular subcomponent of $\Psi_X$ (colored areas) reflects the relative contribution ($[\lambda_g E_g]_i$) of each elementary pulse $g_i(t)$ to the total power.

For mixtures of pulse functions with more complex shapes (e.g. alpha functions), $\Psi_X$ once more corresponds to the linear superposition of the pulses at play (fig. \ref{fig:pulseMix}E-H) --- aligned at $t^\prime = 0$ --- assembled into intricate patterns (from now referred as $\Psi$-patterns) featuring alternating peaks and valleys (e.g. fig. \ref{fig:pulseMix}H). As $\Psi$-patterns enclose an area representing the total power of the signal, the particular segregation of this power can be interpreted as \textit{power density in the time domain}. For example, in fig. \ref{fig:pulseMix}E-G a progressive increase in the power density related to arrhythmic fast phenomena can be observed. Importantly, numerical simulations predict that very fast transients can produce noticeable peaks in $\Psi$-patterns even if their contribution to the total power is small (fig. \ref{fig:pulseMix}H, ultra-fast component).

Finally, it is interesting to analyze a mixture of alpha functions and square pulses  to illustrate the response to mixtures of pulses of different classes. As alpha functions have a maximum at $t = \tau$, the example is constructed mixing alpha functions ($\tau = 100$ ms) with square pulses $100$ ms width (fig. \ref{fig:pulseMix}I). $\Psi_X$ shows the alpha function raised by a constant value from $t^{\prime}=0$ to its maximum and then returning to the alpha profile as expected. 

%---------------------------------------------------------------------------------
\subsection*{Analysing other deviations from a simple homogeneous Poisson process.}
The model analyzed so far considers a purely arrhythmic process (i.e. homogeneous Poisson process). Introducing rhythmic fluctuations to the pulse density $\lambda$ generates a sinusoidal $\Psi_X$ (supplementary fig. \ref{fig:nonHomogPoisson}B). The degree of damping of this sinusoid reflects the bandwidth of $\lambda$ (supplementary fig. \ref{fig:nonHomogPoisson}C,E-F). Given the well-known outcome of the autocovariance function to oscillatory phenomena, this behavior is to be expected. Notice that the oscillatory component in $\Psi_X$ behaves as $sin(t^{\prime})$, also as expected (see eq. \ref{eq:PsiDef}). Remarkably, the pulse is recovered, without alteration, on top of the oscillation produced by the inhomogeneous Poisson process. If the width of the pulse is much smaller than the period of the rhythm, the pulse appears segregated from the sinusoidal fluctuations induced oscillatory $\lambda$ density variations (supplementary fig. \ref{fig:nonHomogPoisson}B-C), otherwise, the pulse and the sinusoidal component may merge (supplementary fig. \ref{fig:nonHomogPoisson}E-F). In this scenario, although a heavily damped sinusoid (associated with a high-bandwidth rhythm) is difficult to identify (supplementary fig. \ref{fig:nonHomogPoisson}F), the magnitude of the first negative phase of $\Psi$ can be considered a good indication of an inhomogeneous Poisson process, as $\Psi$ of a pure homogeneous Poisson process do not exhibit negative phases (supplementary fig. \ref{fig:nonHomogPoisson}A,D).

The simple stochastic process investigated up to this point involves a superposition of equal-amplitude pulses. Richer pulse amplitude distributions in simulations designed to replicate situations such as the effect of electrode distance or a mixture of graded depolarizing and hyperpolarizing potentials demonstrate the robustness of $\Psi$ analysis because the recovery of the underlying pulse is unaffected in these cases (see supplementary fig. \ref{fig:scaledImpulses}).

\begin{figure}[H]
  \centering
    \includegraphics[width=12.5cm, keepaspectratio]{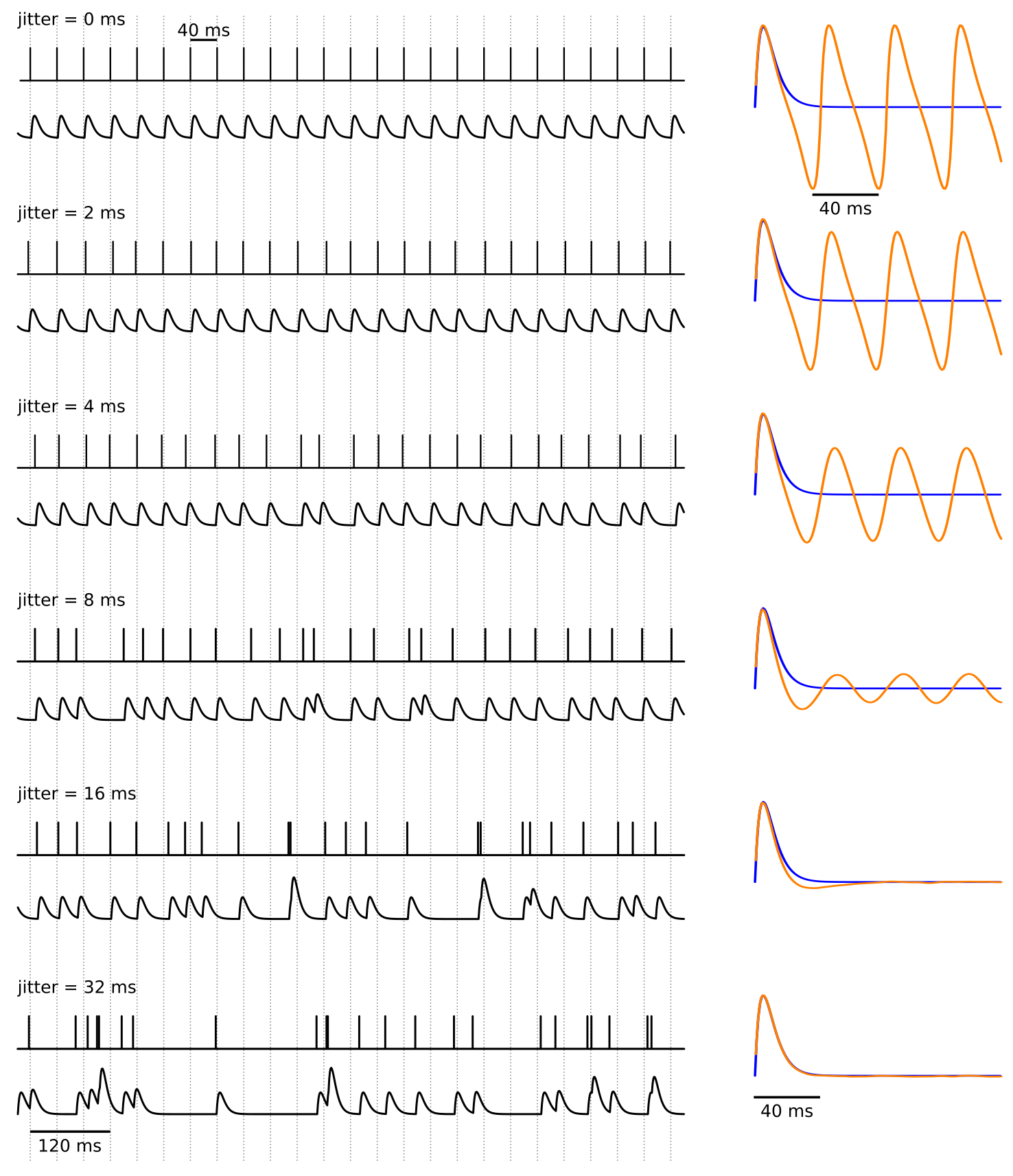}
  \caption{\linespread{1.2} \footnotesize\textbf{(Supplementary) $\boldsymbol{\Psi}$-patterns of quasiperiodic pulse trains modulated by jitter.} Analysis of trains of pulses (alpha function $\tau = 0.005$ ms) produced around a precise period ($40$ ms) distorted by a random Gaussian jitter --- from top to bottom $0, 2, 4, 8, 16, 32$ ms. The left panels show 1-second traces of the impulse functions (top) with the resulting pulse train (bottom) after convolution with the alpha function (see methods). A time reference (spaced at $40$ ms)  is represented by dotted vertical lines. $\Psi$-analysis was performed over $1$-hour pulse trains (@ $2000$ samples per second). The recovered $\Psi$-pattern (orange) is displayed in the right panels alongside the underlying pulse (blue). It is important to note that the recovered pattern may be separated into two distinct components: an initial pulse with constant amplitude regardless of jitter value, and subsequent periodic replicas of this pulse that fade away when rhythmicity is destroyed by growing jitter. The pulse train is statistically indistinguishable from a homogeneous Poisson process when the jitter is significant (e.g., $32$ ms).}   
  \label{fig:jitter}
\end{figure}

Lastly, $\Psi$-analysis performed on a pulse train whose periodicity is disrupted by jitter yields an intriguing result. The jitter in the simulation (supplementary fig. \ref{fig:jitter}) controls a gradient of rhythmicity ranging from perfectly rhythmic to indistinguishable from a Poisson process. All corresponding $\Psi$-patterns have a first pulse component (starting at $t^{\prime} = 0$) with a constant amplitude regardless of the rhythmicity of the input signal and a periodic tail whose amplitude decreases as the rhythmicity of the input signal fades.

%---------------------------------------------------------------------------------
\subsection*{EEG $\boldsymbol{\Psi}$-patterns.}
Let us consider a representative one-hour EEG segment and apply to it the $\Psi$ operator. Figure \ref{fig:EEG_analysis}A shows the corresponding hypnogram depicting the sequence of the major behavioral states: NREM sleep, REM sleep, and wakefulness. The well-known spectral profiles for these states are shown in fig. \ref{fig:EEG_analysis}B (representative of $40$-second periods of the corresponding states) and their dynamic changes (time-frequency analysis) consolidated in the spectrogram provide a classic portrait of sleep-wake dynamics (fig. \ref{fig:EEG_analysis}C). Applying the $\Psi$ operator to EEG ($\Psi_{EEG}$) reveals unique $\Psi$-patterns associated with behavioral states (fig. \ref{fig:EEG_analysis}D). Importantly, these patterns were obtained from EEG sampled at $5$ kHz as they are related to fast phenomena (sampling rates lower than $500$ Hz do not produce any new valuable information).

In addition, paralleling the spectrogram, a time sequence of consecutive $\Psi_{EEG}$ patterns (epoch-based analysis) can be visualized as a heat-map (fig. \ref{fig:EEG_analysis}E) where $\Psi_{EEG}$ profiles are color coded. As stated before, $\Psi$ analysis represents particular arrangements of signal power (i.e. power density). In this regard, an interesting difference between $\Psi_{EEG}$ patterns and EEG spectra is how the corresponding power is \textit{localized}. For example, it is well-known that EEG spectral analysis shows an excess of power distributed all over the beta and gamma bands ($\sim 20$-$80$ Hz) during wakefulness and REM sleep (fig. \ref{fig:EEG_analysis}A,C). 

In parallel, during the same states $\Psi_{EEG}$ exhibits a distinctive narrow peak with maximum at $\sim 3$ ms (fig. \ref{fig:EEG_analysis}D,E labels \textcircled{$1$}, \textcircled{$6$}). Furthermore, there is an even narrower peak ($< 1$ ms width) that appears only during wakefulness (fig. \ref{fig:EEG_analysis}D label \textcircled{$\ast$}, this special feature is shown in detail in fig. \ref{fig:UFC}). On the other hand, while the frequency domain reveals the theta rhythm (observable during wakefulness and REM sleep) as a well-localized peak, $\Psi$ analysis shows a damped oscillation (apex and nadir labeled \textcircled{$2$},\textcircled{$3$},\textcircled{$4$} in the wakefulness $\Psi_{EEG}$ pattern, and \textcircled{$8$},\textcircled{$9$} in the REM sleep $\Psi_{EEG}$ pattern, fig. \ref{fig:EEG_analysis}D,E) expectedly similar to the well-known autocovariance profile for periodic signals (from eq. \ref{eq:PsiDef}). In addition, the wide spectral hump observed in the high-gamma band ($\sim 120-160$ Hz) during REM sleep (fig. \ref{fig:EEG_analysis}A,C) has a correlate as a brief ripple in the corresponding $\Psi$ pattern (fig. \ref{fig:EEG_analysis}D,E label \textcircled{$7$}). Finally, during NREM sleep $\Psi$-patterns correlate with the prominent spectral delta band associated with a wider package of energy (fig. \ref{fig:EEG_analysis}D,E label \textcircled{$5$}) a feature further explored in fig. \ref{fig:SD}).

Throughout the wake-sleep cycle, the alternation of these $\Psi_{EEG}$-patterns and their close relationship with behavioral states is  stable. Furthermore, the patterns depicted in fig. \ref{fig:EEG_analysis} are representative of the $\Psi_{EEG}$-patterns observed in over $2500$ hours of EEG data from $19$ animals (see table \ref{tab:animals} in methods).

%---------------------------------------------------------------------------------
\subsection*{EEG $\boldsymbol{\Psi}$-patterns and brain states.}
The capability of $\Psi_{EEG}$ patterns to reveal valuable information about  \textit{brain states} is illustrated in fig. \ref{fig:PCA}. For easy visualization, a three-dimensional subspace of $\Psi_{EEG}$ can be generated using principal component analysis (PCA) (fig. \ref{fig:PCA}B). In the example, only the first $15$ ms of the $\Psi_{EEG}$ patterns (fig. \ref{fig:PCA}A) were used (see methods), emphasizing the idea that $\Psi$ analysis captures the dynamics of fast phenomena tightly linked to brain function and brain states. PCA on $\Psi_{EEG}$ patterns longer than $15$ ms does not improve clustering quality (not shown). Fig. \ref{fig:PCA}B shows two rotations of the 3D subspace based on the first 3 principal components. In $\Psi_{EEG}$ subspace the behavioral states aggregate in well-defined segregated clusters. It is worth noticing that the achieved discrimination is driven only by EEG information. In contrast, classic spectral analysis cannot emulate such results, and EMG information is traditionally needed to overcome this shortcoming (fig. \ref{fig:PCA}C).

%---------------------------------------------------------------------------------
\subsection*{The ultra-fast component of EEG $\boldsymbol{\Psi}$-patterns is wakefulness-specific.}
The very early phase of $\Psi_{EEG}$ patterns has a close relationship with wakefulness, and fig. \ref{fig:UFC} shows this relation. A zoom-in for the first $10$ ms of delay in the $\Psi_{EEG}$ dynamics color map is shown in fig. \ref{fig:UFC}D. It can be observed that the first samples in the delay axis ($\sim 2-3$ samples) reach high values (cyan) during wakefulness and remain low (dark) during the other behavioral states. This ultra-fast component of $\Psi_{EEG}$ (UFC($\Psi$)) has a rich behavior as it can be observed that even within wakefulness it exhibits large variability, better handled at a logarithmic scale (fig. \ref{fig:UFC}E). Notice that UFC($\Psi$) has a trimodal distribution (fig. \ref{fig:UFC}G) and the two upper subpopulations are associated with wakefulness. UFC($\Psi$) variations, within wakefulness,  reflect different behaviors. The highest values of UFC($\Psi$) are mostly associated with grooming and feeding behaviors while intermediate UFC($\Psi$) values are most likely reflecting active wakefulness (supplementary fig. \ref{fig:actigraphy}).  

The strong relation between UFC($\Psi$) and wakefulness can be also seen in its co-variations with EMG power (fig. \ref{fig:UFC}F). In general, these two variables stay high during wakefulness. Furthermore, in NREM-rich periods UFC($\Psi$) spikes corresponding to micro-arousals can be found. However, despite the global similarities between UFC($\Psi$) and EMG, non-linearities are revealed by the complex clustering  exhibited by the scatter plot between these two variables (fig. \ref{fig:UFC}I). In addition, the histogram of UFC($\Psi$) shows clean segregation along three subpopulations (fig. \ref{fig:UFC}G).  Very low values of UFC($\Psi$) correspond mostly to sleep (left side from arrowhead) while intermediate and high UFC($\Psi$) values are almost exclusively associated with wakefulness (right side from arrowhead).  In this sense, the variable UFC($\Psi$) is, like the classic use of EMG (fig. \ref{fig:UFC}H), a clear indicator of wakefulness but with a more strict separation (compare the narrowness pointed out by arrowheads in both histograms). As aforementioned, the ultra-fast peak associated with UFC($\Psi$) has $\sim 2-3$ samples at 5kHz. Increasing the temporal resolution to 12.5kHz, a clear spike structure can be observed featuring a raising phase, maximum, and falling phase (fig. \ref{fig:UFC}J).

%---------------------------------------------------------------------------------
\subsection*{The slow component of EEG $\boldsymbol{\Psi}$-patterns correlates with delta power.}
\begin{figure}[H]
  \centering
    \includegraphics[width=16cm, keepaspectratio]{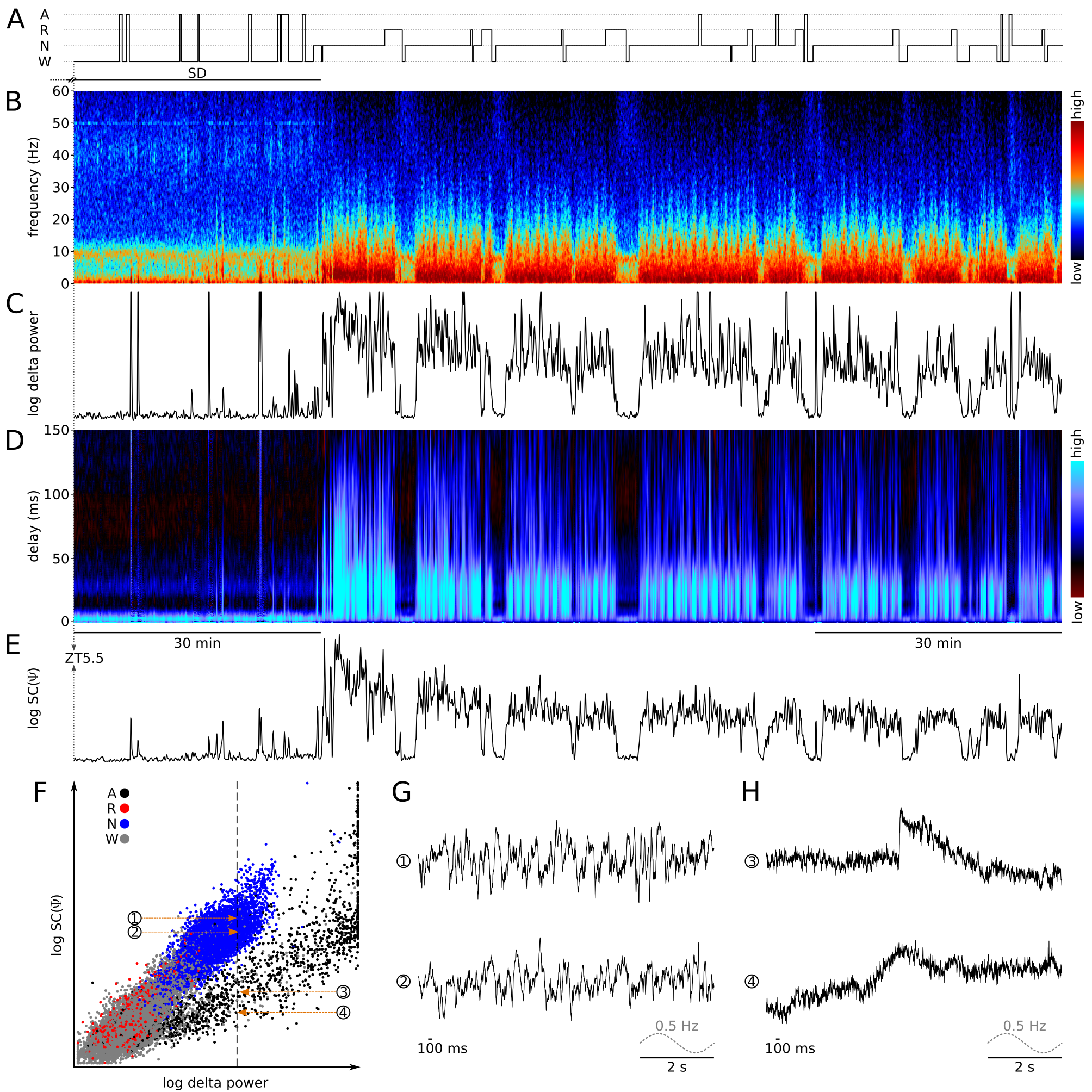}
    \caption{\linespread{1.2} \footnotesize\textbf{Slow component (SC) of $\boldsymbol{\Psi_{EEG}}$ patterns.} }
  \label{fig:SD}
\end{figure}

EEG delta power is the main spectral feature associated with sleep physiology \citep{Borbely1981-xe, Brunner1993-pi}. As shown above, when delta power is high, the corresponding $\Psi_{EEG}$ patterns exhibit the shape of a relatively slow pulse (fig. \ref{fig:EEG_analysis}D, blue trace). From this parallel, it is possible to devise a variable based on $\Psi$ patterns that can serve as a counterpart to EEG delta power. With this goal, we defined the slow component of $\Psi_{EEG}$ (SC($\Psi$)) integrating the power of the first $100$ ms of $\Psi_{EEG}$ patterns (see methods for details). In fig. \ref{fig:SD}, the properties of SC($\Psi$) are explored by comparing it with delta power in the context of 6-hour sleep deprivation (SD) experiments. The last 30 min of the SD protocol together with the following 1.5 hours of recovery sleep are shown. Notice that during the SD protocol, a weak $50$ Hz band can be seen in the spectrogram as the animal cage is open to allow access for gentle handling procedure (fig. \ref{fig:SD}B). The expected delta power rebound \citep{Franken1991-pd, Greene2004-mp} can be observed immediately following the end of SD (fig. \ref{fig:SD}C). SC($\Psi$) is shown in fig. \ref{fig:SD}E, mirroring the dynamics of delta power. The scatter plot SC($\Psi$) vs delta power (log scale) shows a main linear organization of the epochs classified as the three major behavioral states (blue, red, and grey dots) and a secondary branch mostly made up by artifact epochs induced by animal handling (black dots). 

EEG movement-induced artifacts are common in SD experiments and should be tagged because they carry spurious delta power. The observed bifurcation in the scatter plot reveals that SC($\Psi$) is considerably less sensitive to spurious delta power. For example, the dashed line in fig. \ref{fig:SD}F indicates a particular delta power level comprising both genuine delta power and artifacts. Illustrative epochs around this delta power level, taken from the NREM sleep cluster (\textcircled{$1$}, \textcircled{$2$}; blue epochs) and from the movement artifact branch (\textcircled{$3$}, \textcircled{$4$}; black epochs) are presented in \ref{fig:SD}G,H. The conspicuous morphological differences between NREM sleep epochs (\textcircled{$1$}, \textcircled{$2$}) and the artifact epochs (\textcircled{$3$}, \textcircled{$4$}) are well distinguished by SC($\Psi$) as they are projected in significantly different ranges onto the SC($\Psi$) axis. While delta band calculations use the information located within $500-2000$ ms windows, on the other hand, SC($\Psi$) uses time windows up to 20 times shorter ($100$ ms).  The longer time windows needed to calculate delta power are more vulnerable to gathering spurious non-biological phenomena. To help  visualize these temporal relations fig. \ref{fig:SD}G,H have two temporal scales: $100$ ms and $2$ s, together with a sinusoid of 2 second period (i.e. $0.5$ Hz) corresponding to the lower limit of delta band definition.
\end{multicols}
%---------------------------------------------
%\newpage
\noindent\rule{\textwidth}{0.4pt}
\begin{multicols}{2}
\section*{Discussion}
Extracellular neural signals (EEG, LFPs) arise from the superposition of the electrical activity of functionally heterogeneous neuronal populations arranged in topologically complex networks \citep{Bullmore2009-ak}. Classic interpretations of EEG/LFPs favor simplifications of the underlying complexity of neuronal dynamics, assuming that these signals reflect the collective behavior of synchronized oscillatory neuronal populations \citep{Buzsaki2012-ev}. Accordingly, neuronal activity has been conceptualized as being produced by an arrangement of \textit{neural oscillators} which are revealed by the power of particular spectral lines in the signals' frequency domain \citep[][chap.~3]{Niedermeyer6thEd}. Taking into account the overall properties of the EEG spectrum, however, the shortcomings of the conventional interpretation can be revealed by apparent paradoxes, such as the implausibility of imagining the observed EEG $1/f$-type spectrum arising from a perfect line-up of \textit{neuronal oscillators} whose power is inversely proportional to their frequency \citep[see][box 1]{He2014-zk}. This unlikely scenario falls apart, however, if we assume that the EEG generators are arrhythmic pulses. This alternative hypothesis does not associate specific spectral lines with \textit{neuronal oscillators} because these lines as a whole reflect the multi-spectral signature of the generating pulses. Indeed, several lines of evidence support the notion that neuronal dynamics are arrhythmic/stochastic in nature \citep{motokawa1943, elul1969gaussian, Bullock2003-wl, Diaz2007-gk, Freeman2009-rx, He2010-zf, He2014-zk, Cole2017-om, Diaz2018-mu}.

\subsection*{Statistical recovery of a FPP's underlying pulse(s).}
The present study addresses the generation of EEG from a statistical perspective, assuming an underlying stochastic process of pulse superposition. The model (eq. \ref{eq:model}) follows the superposition principle (i.e. pulses add linearly) which is a common assumption supported in the quasi-static approximation of Maxwell's equations \citep{Nunez2006, Linden2013-qb}. At a critical pulse density, the simulated superposition ($X$) aggregates to colored Gaussian noise (fig. \ref{fig:model}A-C). Importantly, this result shows how a $1/f$-like spectrum emerges from the stochastic interference of random pulses (figs. \ref{fig:model}B and \ref{fig:SpecTilt}). 

The study of the mathematical object $\hat{X}$, obtained by subtracting from $X$ delayed versions of itself (eq. \ref{eq:Xhat}), shows how power is injected in the original signal $X$ (fig. \ref{fig:model}D,F). Importantly, the variance of $\hat{X}$  (eq. \ref{eq:var&autocov}) can be used to obtain the underlying pulse $g(t)$ from the original signal $X$. In effect, simulations of superposition for pulse functions $g(t)$ commonly used in neural modeling (exponential function, alpha function, dual-exponential function) (fig. \ref{fig:pulses}A-C) show, rather surprisingly, that the variance of $\hat{X}$ behaves simply as the integral of pulse $g(t)$ (eq. \ref{eq:fundamentalEq}). This empirical relation can be mathematically proven for these pulses (supplementary material),  and thus eq. \ref{eq:fundamentalEq} allows a simple way to statistically recover the shape of pulse $g(t)$ from its own stochastic interference by the differentiation of $\sigma^2_{\hat{X}}(t^{\prime})$ (eq. \ref{eq:infPulse}). 

Importantly to EEG analysis, pulse functions portraying neuronal dynamics do satisfy eq. \ref{eq:fundamentalEq} (fig. \ref{fig:pulses}A-C). However, in general, even though not all $g(t)$ satisfy eq. \ref{eq:fundamentalEq}, the differentiation of $\sigma^2_{\hat{X}}(t^{\prime})$ still provides a good approximation of $g(t)$ given that both functions $h_1(t^\prime)=E_g - (g \star g)(t^\prime)$ and $h_2(t^\prime)=\int_0^{t^{\prime}} g(t)dt$ (see eq. \ref{eq:generalEq}) are similar increasing monotonic functions approaching their asymptotic limits at the same time. Consequently, the differentiation of $h_1(t^\prime)$ generates a pulse with at least the same width as $g(t)$ (fig. \ref{fig:pulses}E-H). 

The operator $\Psi$ applied to a stochastic superposition of some arbitrary pulse $g(t)$ is able to recover an approximation of the underlying pulse $g(t)$ (fig. \ref{fig:model}H and fig. \ref{fig:pulses}). Although $\Psi$ is a transformation within the time domain, the delay ($t^{\prime}$) has a crucial twist as it corresponds to \textit{lags} with respect to \textit{any time} in the original signal $X(t)$. The pulse recovery method embodied by $\Psi$ can be understood as a rearrangement of the interfering pulses composing $X$, so that they end up aligned at $t^{\prime} = 0$. 

Strikingly, $\Psi_X$ eliminates the stochastic component while revealing $g(t)$ thus deconvolving the Filtered Poison Processes $X(t)$. This emergent order is also apparent when applying $\Psi_X$ to signals originating in mixtures of different kinds of pulses, aligning the underlying components at $t^{\prime} = 0$ into unique patterns (fig. \ref{fig:pulseMix}). These $\Psi$-patterns can be interpreted as \textit{power density in the time domain}, as the integral of $\Psi_X$ corresponds to $\sigma^2_X$, related to the signal power (eq. \ref{eq:var&Pow}). Hence, $\Psi$-patterns work analogously to power spectra, describing particular inhomogeneous power segregation in the time domain.

\subsection*{Application of $\boldsymbol{\Psi}$ to EEG}
The temporal dynamics of $\Psi_{\mathit{EEG}}$-patterns can be visualized (fig. \ref{fig:EEG_analysis}E) as a 2D color-coded representation ([\textit{time, delay}]) highly correlating with behavioral states (fig. \ref{fig:EEG_analysis}A). This representation of brain dynamics is analogous to the classic EEG spectrogram (fig. \ref{fig:EEG_analysis}C). The main advantage of $\Psi_{\mathit{EEG}}$ is materialized in how different behavioral states are characterized by specific $\Psi_{\mathit{EEG}}$-patterns (fig. \ref{fig:EEG_analysis}D) while spectrograms produce ambiguous features-to-state relations (fig. \ref{fig:EEG_analysis}C). For example, the EEG spectral similarities during wakefulness and REM sleep --- for this reason also referred to as paradoxical sleep \citep{Jones2018-mp} --- establish well-known difficulties in the discrimination of these two states, a limitation that is partly overcome by the use of EMG in traditional methods. While the spectral power characterizing wakefulness and REM sleep is broadly scattered over the beta and gamma bands, $\Psi_{\mathit{EEG}}$-patterns show the same power distributed in different temporal profiles featuring narrow peaks (fig. \ref{fig:EEG_analysis}D,E labels \textcircled{$\ast$}, \textcircled{$1$}, \textcircled{$6$}). It is especially remarkable that high-frequency ripples associated with REM sleep (fig.\ref{fig:EEG_analysis}D,E label \textcircled{$7$}) can be revealed from EEG using $\Psi_{\mathit{EEG}}$-patterns. High-frequency ripples --- e.g. hippocampal sharp wave ripples --- have been only reported from LFP recordings after the use of filters \citep{Buzsaki2015} and their properties are characterized by averaging across many instances of properly aligned ripples \citep{Cheah2019-ra}. On the other hand, with $\Psi$ analysis, the observed ripples are a straightforward statistical property of EEG epochs during REM sleep.

Altogether, the constellation of morphological details characterizing $\Psi_{\mathit{EEG}}(t^{\prime})$-patterns (fig.\ref{fig:EEG_analysis}D,E) allows for clear behavioral state segregation in the $\Psi_{\mathit{EEG}}$-space. Remarkably, in addition to favored composed directions in $\Psi_{\mathit{EEG}}$-space (as shown using PCA), selected raw dimensions of the $\Psi_{\mathit{EEG}}$-vectors directly allow a discrimination of behavioral states without further calculations. As shown in supplementary fig. \ref{fig:PCA2}, using the peak of the ultra-fast component (fig. \ref{fig:EEG_analysis}D,E label \textcircled{$\ast$}), the peak of the fast component (fig. \ref{fig:EEG_analysis}D,E labels \textcircled{$1$}, \textcircled{$6$}), and a value representing the maxima of the prominent slow component (fig. \ref{fig:EEG_analysis}D,E label \textcircled{$5$}) can separate  behavioral states. It is worth noting that, due to well-known problems with frequency estimation \citep{Prerau2017-ov}, using specific spectral lines as meaningful features is impractical given the high variability of Fourier spectra. Thus, the result in supplementary fig. \ref{fig:PCA2} is an illustration of the exceptional stability and reliability of $\Psi_{\mathit{EEG}}$-patterns.

The ultra-fast component of $\Psi_{\mathit{EEG}}$ (fig. \ref{fig:EEG_analysis}D label \textcircled{$\ast$} and fig. \ref{fig:UFC}) is the first EEG feature specific for wakefulness reported so far and may offer a promising contribution in the study of sleep-wake dynamics which currently depend largely on EMG to determine wakefulness and arousal. However, EMG has some technical disadvantages compared with EEG as muscle electrodes are not fixed in space, which compromises the long-term stability of the muscle signal (fig. \ref{fig:UFC}E-H). The observed non-linearities and clustering patterns between UFC($\Psi$) and EMG suggest they are independent variables (fig. \ref{fig:UFC}I) --- i.e. ultra-fast components are generated in the brain and do not reflect an intrusion of muscle activity by volume conduction. 

Also supporting this interpretation of independent variables, in the moments when UFC($\Psi$) is higher (above quantile $0.75$), the animal is usually engaged in quiet-wake and when the animal is actively exploring, UFC($\Psi$) is  below quantile $0.75$ (supplementary fig. \ref{fig:actigraphy}). It is interesting how the stochastic superposition model can detect narrow $\Psi$ peaks even if they represent a small fraction of the total power (fig. \ref{fig:pulseMix}). On the other hand, in the frequency domain, the energy carried by fast pulses is homogeneously distributed through the spectrum (fig. \ref{fig:model}B). Hence, it is not surprising that the crystal-clear ultra-fast $\Psi_{\mathit{EEG}}$ component does not have a noticeable corresponding feature in the EEG spectrum. 

\begin{figure}[H]
  \centering
    \includegraphics[width=8.5cm, keepaspectratio]{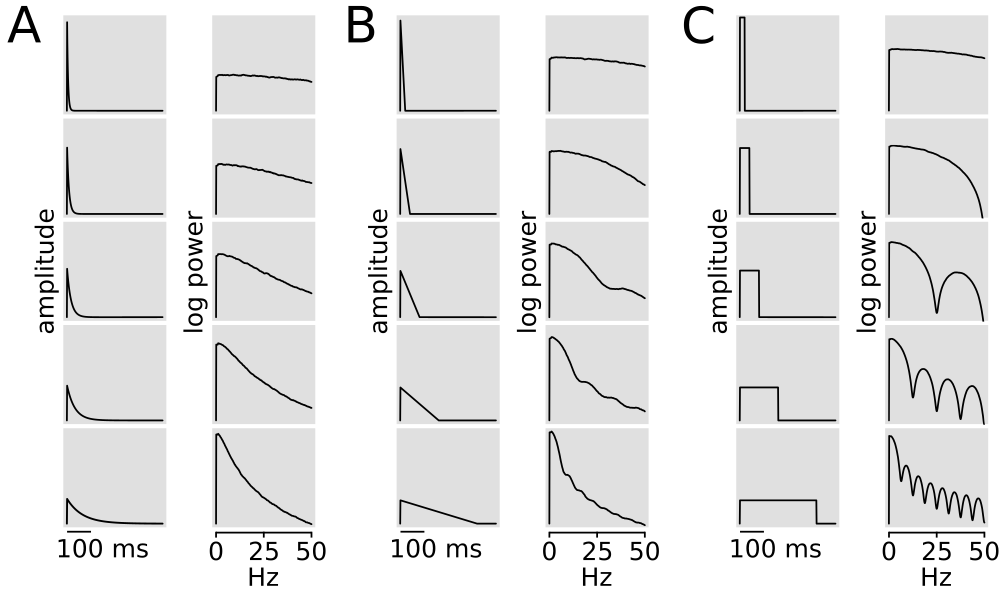}
  \caption{\linespread{1.2} \footnotesize\textbf{(Supplementary) Pulse width and $1/f$-like spectrum.} Figure \ref{fig:model}B shows the relation between a sequence of alpha functions with increasing $\tau$ and the progressive slope of the spectrum of process $X$. Here, an extension of this analysis is presented considering other pulse profiles. \textbf{A.} Exponential decay; $\tau = 2, 4, 8, 16, 32$ ms. \textbf{B.} Triangular pulse (sawtooth); width = $20, 40, 80, 160, 320$ ms. \textbf{C.} Square pulse; width = $20, 40, 80, 160, 320$ ms. In each panel, the left column shows the pulse increasing its width from top to bottom. The amplitude scale is fixed and all pulses have unit energy. The right column shows the corresponding power density spectrum (semi-log) for a superposition of the corresponding pulses ($\lambda_g = 10^4$ pulses per second).}   
  \label{fig:SpecTilt}
\end{figure}

The power of the EEG delta band is the most prominent EEG spectral feature used in the study of sleep-wake cycle dynamics. These dynamics can be alternatively studied  in $\Psi_{\mathit{EEG}}$-space by following the variations of the slow packet of energy shown in  fig. \ref{fig:EEG_analysis}D,E label \textcircled{$5$}. In fact, an evaluation of the energy associated with this energy packet, SC($\Psi$) (see methods) is able to parallel the well-known delta power temporal profile triggered by sleep deprivation (fig. \ref{fig:SD}E). The correlation between SC($\Psi$) and delta power is good when the motion-artifact epochs are discarded (fig. \ref{fig:SD}F). The SC($\Psi$) variable can discriminate  between motion-artifact epochs (fig. \ref{fig:SD}H) and physiologically occurring epochs containing  high EEG delta power (NREM epochs; fig. \ref{fig:SD}G). Discrimination based only on delta power fails as it gets confused by the particular temporal profile of motion artifacts which typically contain high-amplitude slow oscillations. In addition, fig. \ref{fig:SpecTilt} shows that stochastic superpositions composed of wide pulses, regardless of their shapes, have high delta power, revealing delta power to be an unspecific side effect of stochastic superpositions. Furthermore, several pieces of empirical evidence warn about delta power's lack of specificity as a sleep biomarker \citep{Davis2011-he}.

\subsection*{Concluding remarks}
Understanding the brain's functional organization is a complex task as observations from micro (single cell recordings), meso (LFP) and macro (EEG) spatial scales must be conceptually integrated \citep{Deco2008-yn}. Even though the observed phenomenology across these spatial scales emerges from the same general biophysical principles, there is a traditional qualitative gap in the description of the macro-scale, on the one hand, ruled by low-frequency \textit{rhythmic oscillations} \citep{Timofeev2011, Adamantidis2019-ed}, while on the other hand, the micro- and meso-scales seem driven by fast transient phenomena \citep{Luczak2015-yq}. 

\begin{figure}[H]
  \centering
    \includegraphics[width=16cm, keepaspectratio]{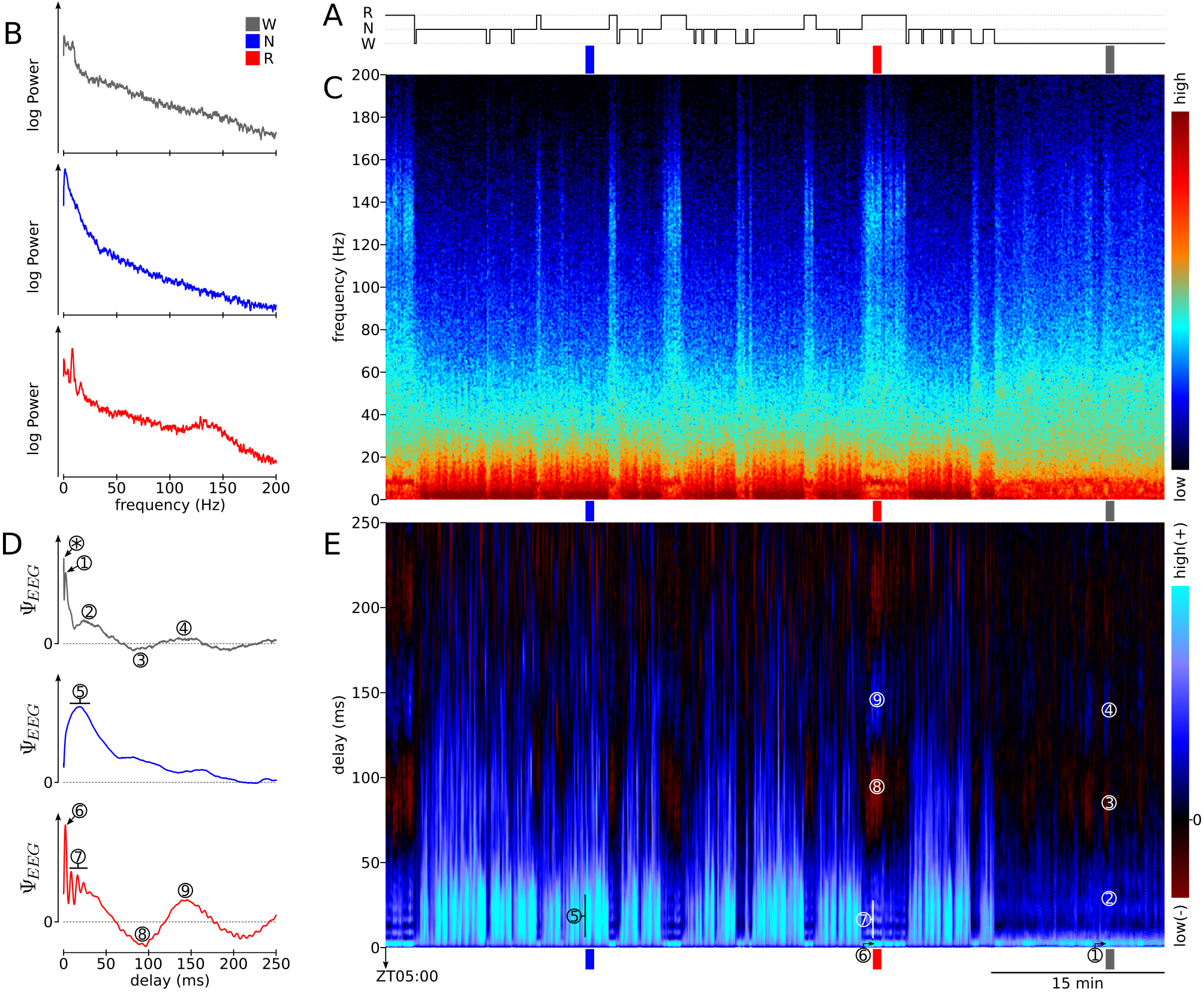}
    \caption{\linespread{1.2} \footnotesize\textbf{$\boldsymbol{\Psi_{EEG}}$ patterns.} \textbf{A.} Hypnogram of a representative one-hour spontaneous mouse EEG (M3, table \ref{tab:animals} in methods) starting at ZT 5 (R = REM sleep, N = non-REM sleep, W = wakefulness). \textbf{B.} Three spectra representative of wakefulness (gray), non-REM sleep (blue), and REM sleep (red), corresponding to the 40-second segments (10 epochs) highlighted by rectangles similarly colored in panels A, C, and E. \textbf{C.} Multitaper spectrogram of the one-hour spontaneous EEG calculated at a 4-second epoch resolution. Pseudo-color represents log-spectral power (black $<$ blue $<$ cyan $<$ orange $<$ red $<$ dark-red). \textbf{D.} $\Psi_{\mathit{EEG}}$ of the same representative segments shown in panel A. Encircled labels (\textcircled{$1$}, \textcircled{$2$}... \textcircled{n}) highlights special features of $\Psi_{\mathit{EEG}}$ patterns (see main text). \textbf{E.} Temporal dynamics of $\Psi_{\mathit{EEG}}$ patterns visualized as a $[\mathit{time}, \mathit{delay}]$ space. For the original one-hour EEG period, $\Psi_{EEG}$ patterns were calculated at a resolution of 4-second epoch (see methods) and plotted as vertical lines color-coded according to the pattern value. Pseudo-color scale: red $<$ black = 0 $<$ blue $<$ light-blue $<$ cyan.  Encircled labels point to the same features as in panel D.} 
  \label{fig:EEG_analysis}
\end{figure}

Our findings show fast transient phenomena in EEG --- i.e. $\Psi_{EEG}$ patterns with temporal properties akin to those observed at the micro- and meso-scales ---  providing a missing link to understand neuronal dynamics at any scale. Based on our stochastic pulse mixture model (fig. \ref{fig:pulseMix}) the empirical  $\Psi_{EEG}$-patterns obtained (fig. \ref{fig:EEG_analysis}) can be interpreted as the superposition of different transients of neuronal activity --- aligned at $t^{\prime} = 0$ --- weighted by their respective energy contribution. The neuronal transients involved should match the well-known events at the  basis of the biophysical neural computation occurring at different time scales, such as axonic action potentials ($\sim1$ ms), dendritic Na\textsuperscript{+} spikes ($\sim 1$ ms), Ca\textsuperscript{2+} spikes ($\sim 10-100$), NMDA spikes ($\sim 50-100$ ms), fast EPSPs (e.g. AMPA; $\tau \sim 2-3$ ms), slow EPSPs (e.g. NMDA; $\tau > 80$ ms), among others \citep{Durstewitz2009-jw, Buzsaki2012-ev, Larkum2022-lc}. Further research is required to associate $\Psi_{EEG}$ patterns to specific biophysical processes.

Technically, the time scales of $\Psi$-patterns require a sampling rate acquisition of at least $5$ kHz. Although it is of utmost interest to apply $\Psi$ analysis to human EEG, unfortunately, commonly available instrumentation (e.g. clinical grade) is not intended to perform data acquisition at sampling rates as high as those used in this study ---as conventional EEG analysis focuses on \textit{oscillations} lower than $100$ Hz. Our results suggest revising the technical standards for human EEG studies.

Oscillatory phenomena are ubiquitous in nature, and Fourier analysis offers an excellent tool describing their fundamental properties in terms of amplitude, frequency, and phase. However, the versatility of Fourier analysis requires that any periodic signal has an equivalent Fourier series but the frequencies appearing in the series are a mathematical convenience and do not reflect necessarily the existence of physical oscillators. For this reason, alternatives to Fourier analysis are required to gain a broader insight. $\Psi$ analysis provides an interesting characterization for arrhythmic non-stationary phenomena (fig. \ref{fig:pulseMix}), while at the same time giving a proper description of oscillatory phenomena (e.g. theta rhythm and high-frequency ripples; fig. \ref{fig:EEG_analysis}D,E) linked to the well-known properties of the auto-covariance function (eq. \ref{eq:PsiDef}). Furthermore, the stochastic notions that underpin $\Psi$ analysis pave the way for the Chapman-Kolmogorov equation to be applied to EEG dynamics \cite{Eden2004-ps}. Beyond neuronal dynamics, $\Psi$ analysis and its theoretical framework potentially have broad applications in Natural Sciences and Engineering.
\end{multicols}
%---------------------------------------------
\newpage
\section*{Figures}
%---------------------------------------------------------------------------------

\newpage
{\footnotesize
\noindent
\textbf{A.} Realizations  ($X(t)$) of the stochastic superposition by pulse $g(t)$ (alpha function with $\tau = 5 ms$, blue trace).  Black traces illustrate 1-second segments of process $X(t)$. From top to bottom, process $X(t)$ is generated at pulse densities $\lambda_g = 1, 10, 10^2, 10^3$, and $10^4$ pulses per second. All traces share the same amplitude scale. In all superpositions, the DC component is discarded (see methods). \textbf{B.} Spectral properties of $X$ are determined by the pulse width. Left-column (light-ocher) shows a sequence of alpha functions of unitary energy, from top to bottom $\tau =$  $2$, $4$, $8$, $16$, and $32$ milliseconds. Center-column (light-blue) shows 2-second traces of process $X$ generated from the corresponding pulse (same row) at $10^4$ pulses per second. Right-column (pink) shows the corresponding semi-log spectra of $X$. \textbf{C.} Dark traces correspond to $250$ realizations of process $X(t)$ generated with $\lambda_g = 10^4$ pulses per second ($1000$-millisecond traces but only 100 ms are visualized). Red trace corresponds to the time-wise standard deviation across the $10^4$ realizations of $X(t)$ ($\sigma_X$). The right inset corresponds to the histogram of all values in the $10^4$ realizations approaching a Gaussian distribution. \textbf{D.} For a new set of processes $X(t)$ generated in equivalent conditions as in C, the signals are translated to zero at the arbitrary time $t_0$ (vertical dotted line) --- i.e. $X(t) - X(t_0)$ --- (250 realizations plotted). The amplitude of the random process after this translation appears to diffuse from zero in a constrained fashion. The red trace is the time-wise standard deviation calculated over $10^4$ realizations of $X(t) - X(t_0)$, describing the temporal profile of this diffusion (this standard deviation is equivalent to $\sigma_{\hat X}$, due to the ergodicity of $X$, see main text). Note that the asymptotic value of this standard deviation  (the plateau of the red trace) is $\sqrt{2}\sigma_X$. The histogram of the right inset does not include the notch, i.e. is calculated with values outside the 100 ms visualization window. The bottom blue trace is the generating pulse $g(t)$. \textbf{E.} Six iso-energetic pulses (blue traces), from left to right two alpha functions ($\tau = 2.5$ ms, and $\tau = 5$ ms), two square pulses (25 ms and 50 ms in length), and two exponentially decaying pulses ($\tau = 10$ ms, and $\tau = 20$ ms). \textbf{F.} The analysis shown in D is applied to processes $X$ generated by the corresponding pulses $g(t)$  shown in E. Plotted together are 250 realizations of $X(t) - X(t_0)$ (black traces) with $\sigma_{\hat{X}}$ (red trace). \textbf{G.} Dark red traces correspond to variance of $\hat X$ ($\sigma^2_{\hat X}$). \textbf{H.} Blue traces correspond to the same pulses shown in E, and dark-red circles correspond to the numerical derivative of $\sigma^2_{\hat X}(t^{\prime})$ ($\frac{d}{dt^{\prime}} \sigma^2_{\hat X}(t^{\prime})$) properly scaled (eq. \ref{eq:scaling}). Observe that this overall procedure recovers pulses $g(t)$ from Gaussian noise. In panels A-D,F,G, black arrowheads indicate \textit{zero} level.}

\newpage
%---------------------------------------------------------------------------------

\begin{figure}[H]
  \centering
    \includegraphics[width=8.5cm, keepaspectratio]{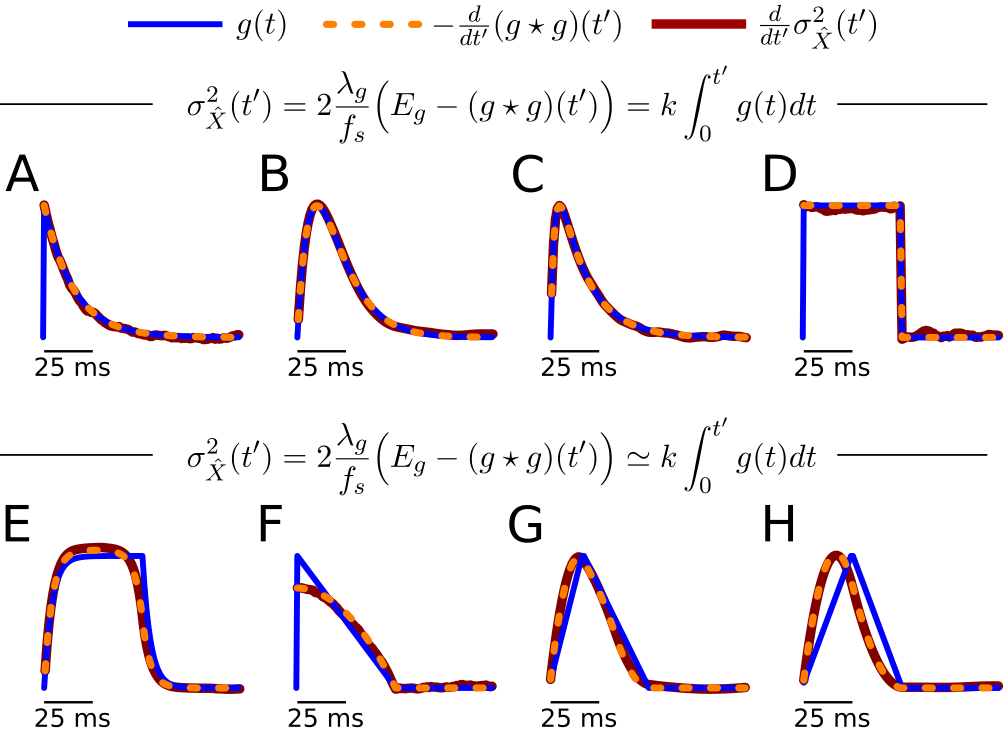}
  \caption{\linespread{1.2} \footnotesize\textbf{Two cases of recovering approximations of $\boldsymbol{g(t)}$ from stochastic process $\boldsymbol{X}$}. As described (fig. \ref{fig:model}A), a sequence of a process $X$ (80-seconds in length at 1000 samples per second) was generated using the pulse shapes $g(t)$ depicted by  blue traces in panels \textbf{A-H} (\textbf{A}, exponential decay ($\tau =$ 15 ms); \textbf{B}, alpha function ($\tau =$ 10 ms); \textbf{C}, dual exponential (raising-phase $\tau =$ 2 ms; decaying-phase $\tau =$ 15 ms); \textbf{C}, square pulse ($50$ ms in length); \textbf{E}, capacitor charge/discharge ($\tau =$ 4 ms; charging phase 50 ms duration); \textbf{F-H} three kinds of triangular pulses (rising-time/falling-time: \textbf{F} $0$ ms/$50$ ms; \textbf{G} $16$ ms/$34$ ms; \textbf{H} $25$ ms/$25$ ms). Orange dashed traces correspond to minus the derivative of the auto-covariance of the pulse $g(t)$ and dark-red thick traces represent the derivative of $\sigma_{\hat{X}}^2$ (recovered pulse). In panels \textbf{A-D} recovered pulses (dark-red trace) match the underlying pulse (blue trace) as these functions satisfy equation \ref{eq:fundamentalEq} (top equation, see main text). Panels \textbf{E-H} show examples of pulses $g(t)$ not satisfying equation \ref{eq:fundamentalEq}, still dark-red and orange dashed-traces match as equation \ref{eq:var&autocov} is general for any $g(t)$ (lower equation). Importantly, panels \textbf{E-H} show that even for pulses not satisfying equation \ref{eq:fundamentalEq}, the recovered pulse is still a good approximation of $g(t)$ as it has the same duration (equation \ref{eq:generalEq}). }
  \label{fig:pulses}
\end{figure}

\newpage
%---------------------------------------------------------------------------------

\newpage
%---------------------------------------------------------------------------------

\newpage
%---------------------------------------------------------------------------------
\begin{figure}[H]
  \centering
    \includegraphics[width=12.75cm, keepaspectratio]{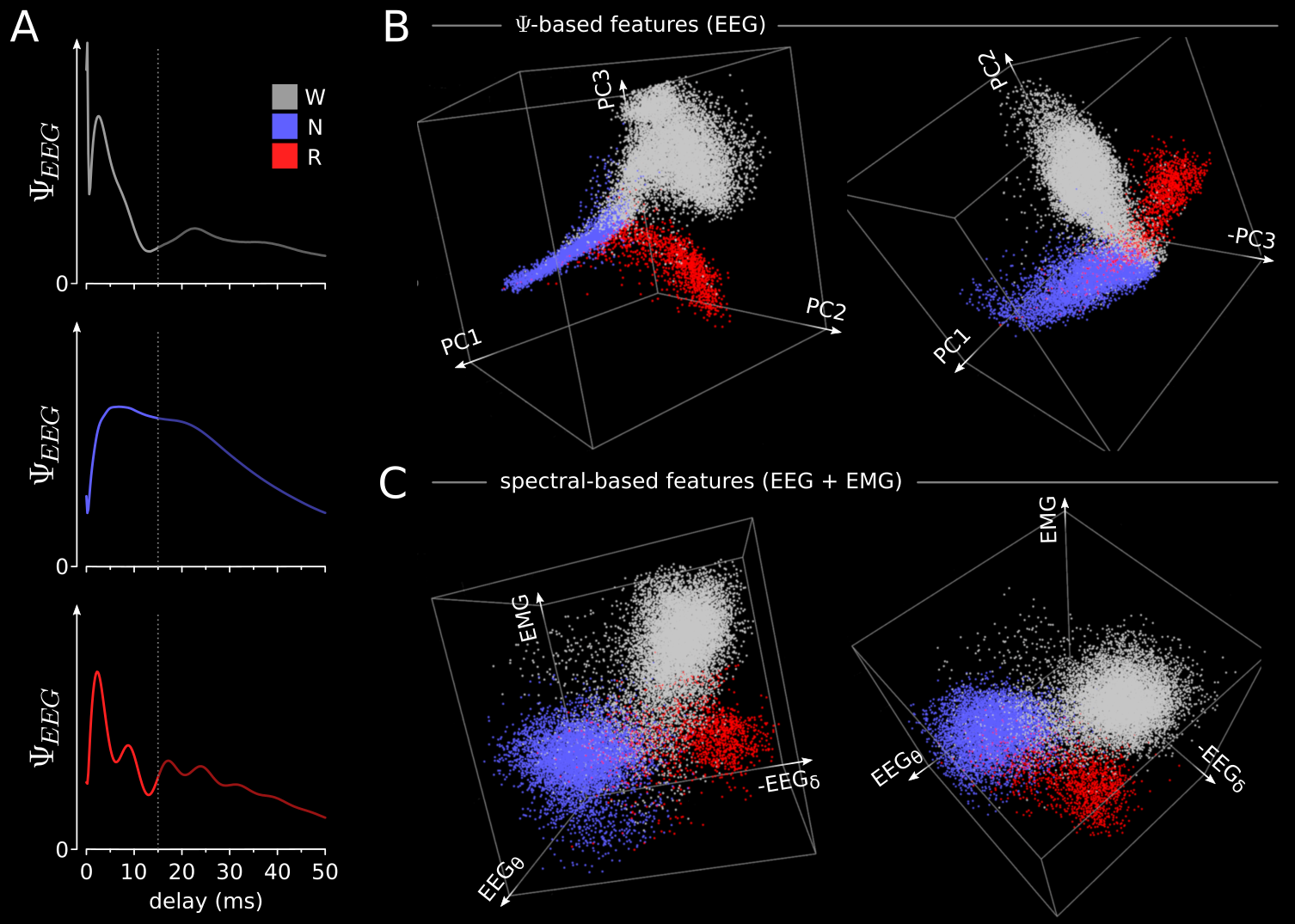}
    \caption{\linespread{1.2} \footnotesize\textbf{Dimensionality reduction of $\boldsymbol{\Psi_{EEG}}$ space.} \textbf{A.} From top to bottom, average $\Psi_{EEG}$ patterns (first 50 ms) for wakefulness (gray), NREM sleep (blue), and REM sleep (red) (average across 24 hours), with their first 15 ms highlighted. \textbf{B.} Using the first 15 ms of $\Psi_{EEG}$ from each 4-second epoch, a standard principal component analysis is performed (see methods), and two rotations of the subspace defined by the first three principal components are presented (PC1-3). \textbf{C.} Classic spectral-based cluster analysis of the same EEG data including its corresponding EMG. The selected features are EEG delta ($\delta$) power ($0.5-4$ Hz), EEG theta ($\theta$) power ($6-12$ Hz), and EMG power ($70-90$ Hz). In panels B and C, each dot corresponds to a 4-second epoch, The analysis is carried out by extending the data shown in figure \ref{fig:EEG_analysis} to the entire 24-hour wake-sleep cycle, or 21600 epochs in total (M3, table \ref{tab:animals} in methods).} 
  \label{fig:PCA}
\end{figure}

\newpage
%---------------------------------------------------------------------------------
\begin{figure}[H]
  \centering
    \includegraphics[width=16cm, keepaspectratio]{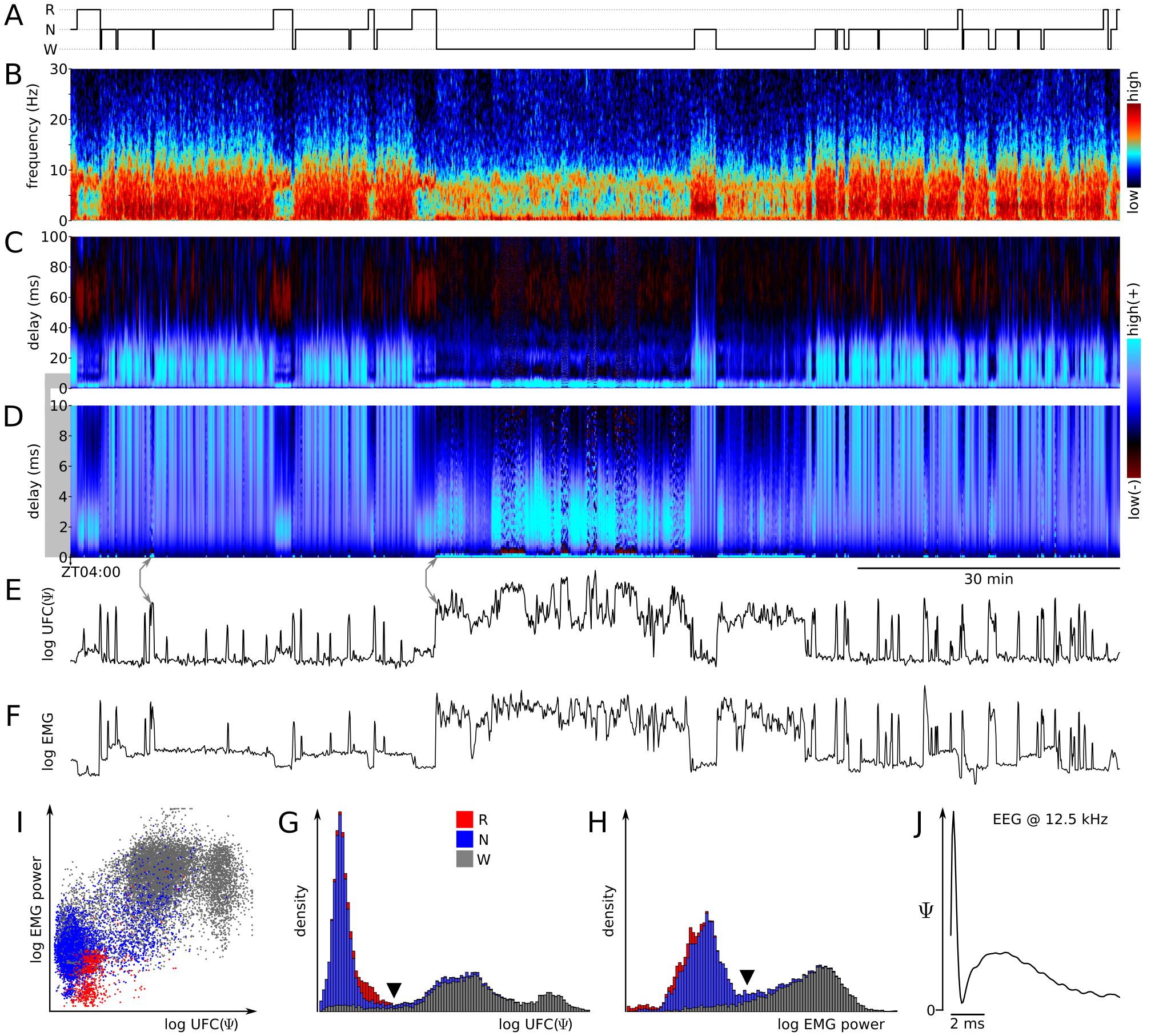}
    \caption{\linespread{1.2} \footnotesize\textbf{Ultra fast component (UFC) of $\boldsymbol{\Psi_{EEG}}$ patterns.} \textbf{A.} hypnogram of a representative two-hour mouse spontaneous EEG (M11, table \ref{tab:animals} in methods). \textbf{B.} Corresponding Multitaper spectrogram. \textbf{C.} Corresponding  $\Psi_{\mathit{EEG}}$ dynamics. \textbf{D.} Zoom in, indicated by the gray shaded area, of the first 10 ms delay ($y$-axis) of the $\Psi_{\mathit{EEG}}$ dynamics shown in panel C. Note that the first pixels (bottom) fluctuate between close to zero values (back) and high-values (cyan). \textbf{E.} Temporal course in logarithmic scale of UFC($\Psi$) (see methods for UFC($\Psi$) definition). Gray arrows indicate examples of the correspondence between high UFC($\Psi$) and the bottom pixels of $\Psi_{\mathit{EEG}}$-dynamics colored map (cyan on panel D). \textbf{F.} The corresponding EMG power in logarithmic scale. \textbf{G.} Histogram of log UFC($\Psi$) (24-hour EEG). The histogram is built as a stacked bar plot, with every bin expressing the proportion of wakefulness (gray), NREM sleep (blue), and REM sleep (red). The black arrowhead indicates the deep density depression separating sleep (NREM + REM) and wakefulness \textbf{H.} Histogram of log EMG power using the same convention as in G. \textbf{I.} scatter-plot log UFC($\Psi$) vs log EMG. Every dot is a $4$-second epoch color-coded according to the behavioral state ($21600$ epochs displayed). \textbf{J.} Example of the ultra-fast spike of $\Psi_{\mathit{EEG}}$ patterns measured during active wakefulness sampled at an enhanced temporal resolution ($12.5$ kHz, M16, table \ref{tab:animals} in methods).} 
  \label{fig:UFC}
\end{figure}

\newpage
%---------------------------------------------------------------------------------

\newpage
{\footnotesize
\noindent
An illustrative portion of a 6-hour sleep deprivation experiment (M18, table \ref{tab:animals} in methods). The visualization (panels A-E), sharing the time scale, encompasses 2 hours of EEG starting at ZT5.5 (i.e. the last 30 minutes of SD are shown followed by 1.5 hours of sleep recovery)  \textbf{A.} Hypnogram (A = epoch with artifacts, R = REM sleep, N = non-REM sleep, W = wakefulness). \textbf{B.} EEG Multitaper spectrogram. \textbf{C.} EEG delta power (log scale). \textbf{D.} $\Psi_{\mathit{EEG}}$-dynamics colored map. \textbf{E.} slow component of $\Psi_{EEG}$ (SC($\Psi$)) in logarithmic scale (see methods for SC($\Psi$) definition). \textbf{F.} Scatterplot of EEG delta power vs SC($\Psi$) (log scale). Numbers indicate 4 selected epochs with approximately the same delta power (vertical dashed line). These epochs are analyzed in panels G and H. \textbf{G.} Raw EEG traces corresponding to two selected epochs (NREM sleep) with delta power close to the level indicated by the dashed line in panel F. \textbf{H.} As in panel G, two artifact epochs have the same delta power level. In panels G, H  the short time scale (100 ms) shows the time window used to calculate SC($\Psi$) and the longer (2 s) serves to compare with a 0.5 Hz sine wave corresponding to the lower limit of the delta band.}

\newpage
%---------------------------------------------------------------------------------

\newpage
%---------------------------------------------------------------------------------
\begin{figure}[H]
  \centering
    \includegraphics[width=8.5cm, keepaspectratio]{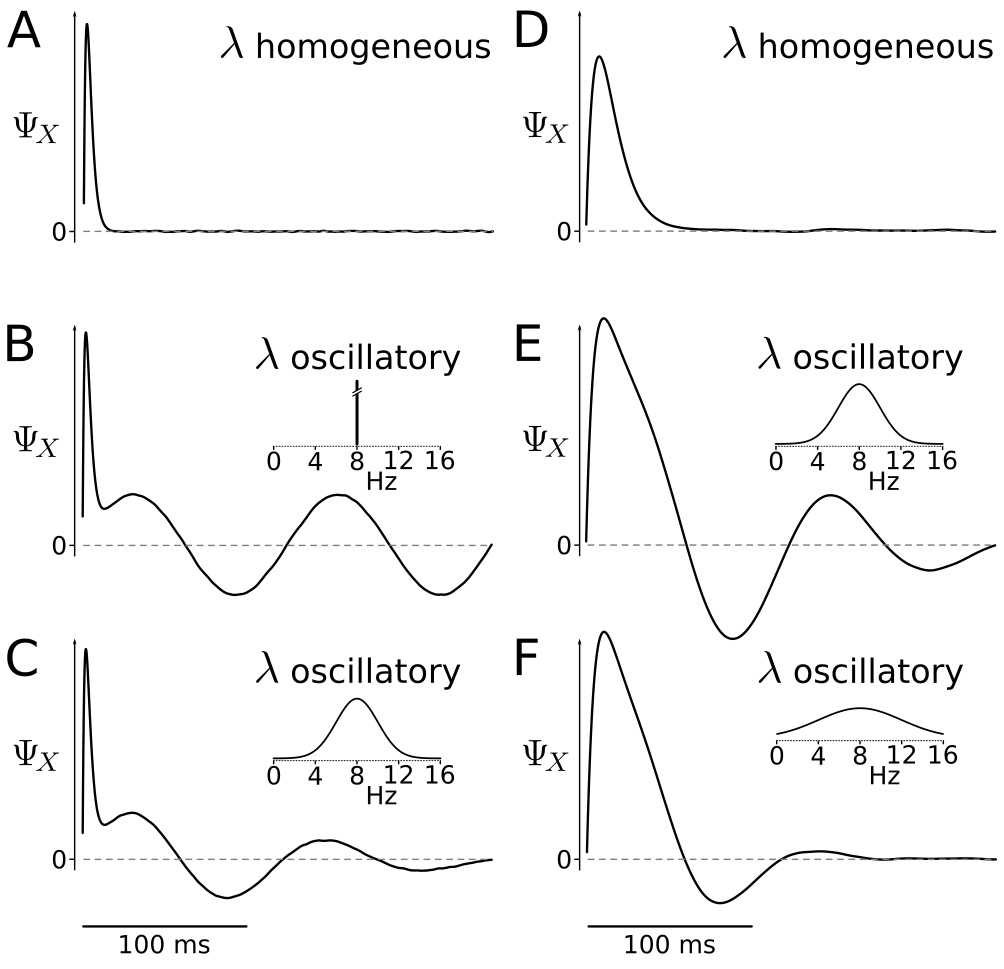}
  \caption{\linespread{1.2} \footnotesize\textbf{(Supplementary) Arrhythmic processes combined with oscillatory phenomena originated from inhomogeneous Poisson processes.} \textbf{A.} $\Psi_X$ for a superposition of alpha functions $\tau = 2$ ms, as in previous simulations the Poisson process is homogeneous (i.e. $\lambda$ is constant). \textbf{B.} $\Psi_X$ for a similar process (as in A) with a pulse density $\lambda$ fluctuating at 8 Hz (see methods for details). \textbf{C.} $\Psi_X$ for a similar process (as in A) with the pulse density $\lambda$ fluctuating at $8 \pm 2$ Hz (Gaussian spectrum). \textbf{D.} $\Psi_X$ for a superposition of alpha functions $\tau = 8$ ms (homogeneous Poisson process). \textbf{E.} $\Psi_X$ for a similar process (as in D) with the pulse density $\lambda$ fluctuating at $8 \pm 2$ Hz. \textbf{F.} $\Psi_X$ for a similar process (as in D) with pulse density $\lambda$ fluctuating at $8 \pm 4$ Hz. Insets show the spectra of $\lambda$. Only panels in the same column (\textbf{A-C}; \textbf{D-F}) share the same vertical scale.}   
  \label{fig:nonHomogPoisson}
\end{figure}

\newpage
%---------------------------------------------------------------------------------
\begin{figure}[H]
  \centering
    \includegraphics[width=11.0cm, keepaspectratio]{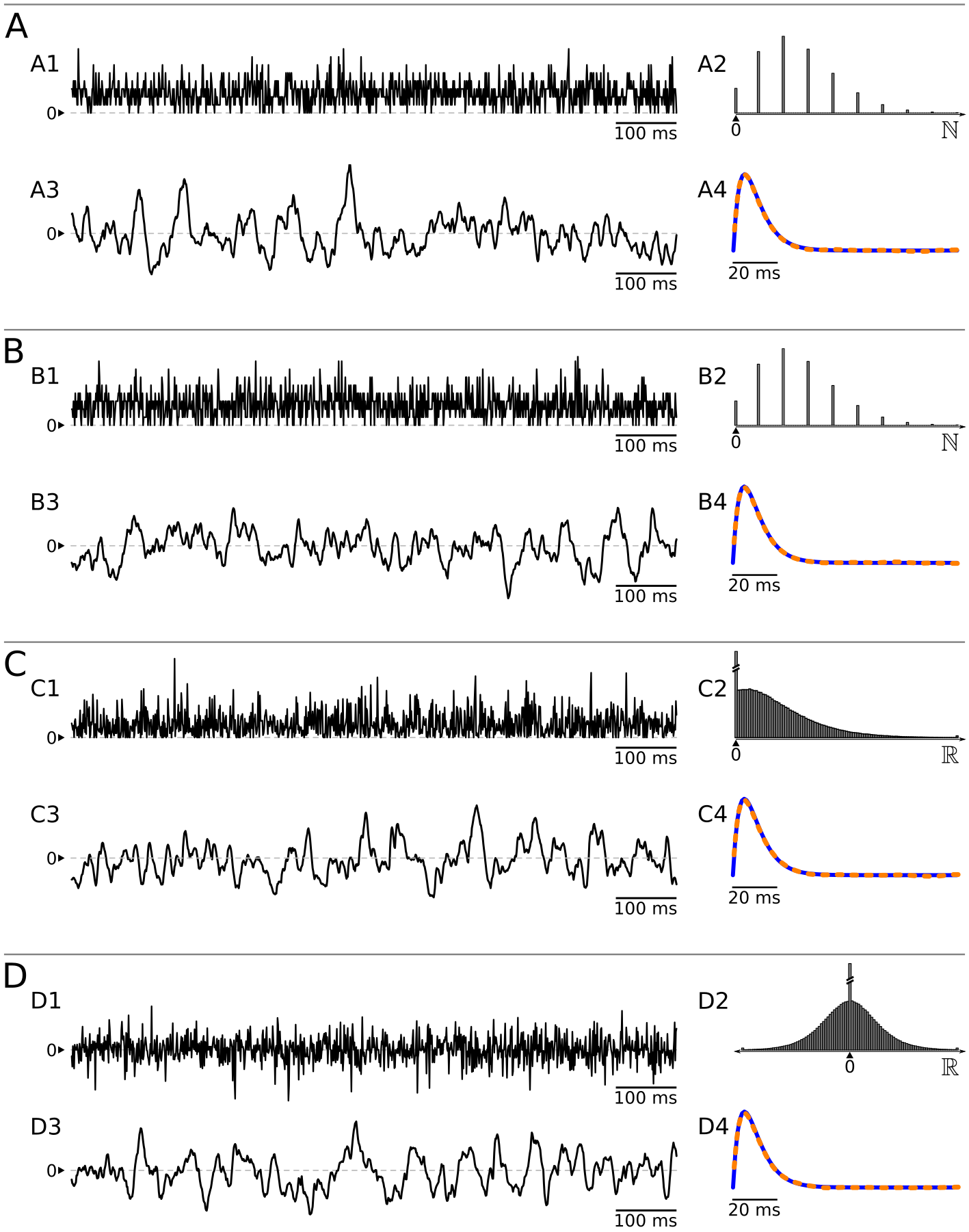}
  \caption{\linespread{1.2} \footnotesize\textbf{(Supplementary) Recovering pulse $\boldsymbol{g(t)}$ from superpositions constructed by different schemes controlling impulse size.} In these examples $g(t)$ is an alpha function ($\tau = 0.005$ ms), the sampling rate is $1000$ samples per second, and the pulse density is $2500$ pulses per second (i.e. $2.5$ pulses per sample).  \textbf{A.} Signal generated by a random superposition of pulses according to a Poisson process as the number of events in each time bin is determined by a random value with a Poisson distribution.  Sub-panel A1 shows a 1-second trace of the number of events per time bin. Sub-panel A2 shows the histogram of the number of events per time bin. Sub-panel A3 shows the convolution of the train of events shown in sub-panel A1 with $g(t)$ producing the actual signal (see methods). Sub-panel 4 shows the recovered pulse from the signal (orange dashed traces alongside the underlying pulse $g(t)$ (blue trace). \textbf{B.} An analogous simulation is conducted with the exception that each unit amplitude impulse is randomly placed in a time bin with a uniform distribution (see methods). This yields a result with identical statistical characteristics as those displayed in panel A. \textbf{C.} Simulation  using the same algorithm as in panel B, but this time each impulse has a random amplitude with exponential distribution. \textbf{D.} Similar to the previous panel, except that each impulse  has a random amplitude with a Gaussian distribution. It is worth noting that $g(t)$ can be recovered regardless of the amplitude distribution of the impulses. Surprisingly, this holds true even for a mixture of positive and negative pulses, as shown by the Gaussian distribution.}   
  \label{fig:scaledImpulses}
\end{figure}

\newpage
%---------------------------------------------------------------------------------

\newpage
%---------------------------------------------------------------------------------
\begin{figure}[H]
  \centering
    \includegraphics[width=8.5cm, keepaspectratio]{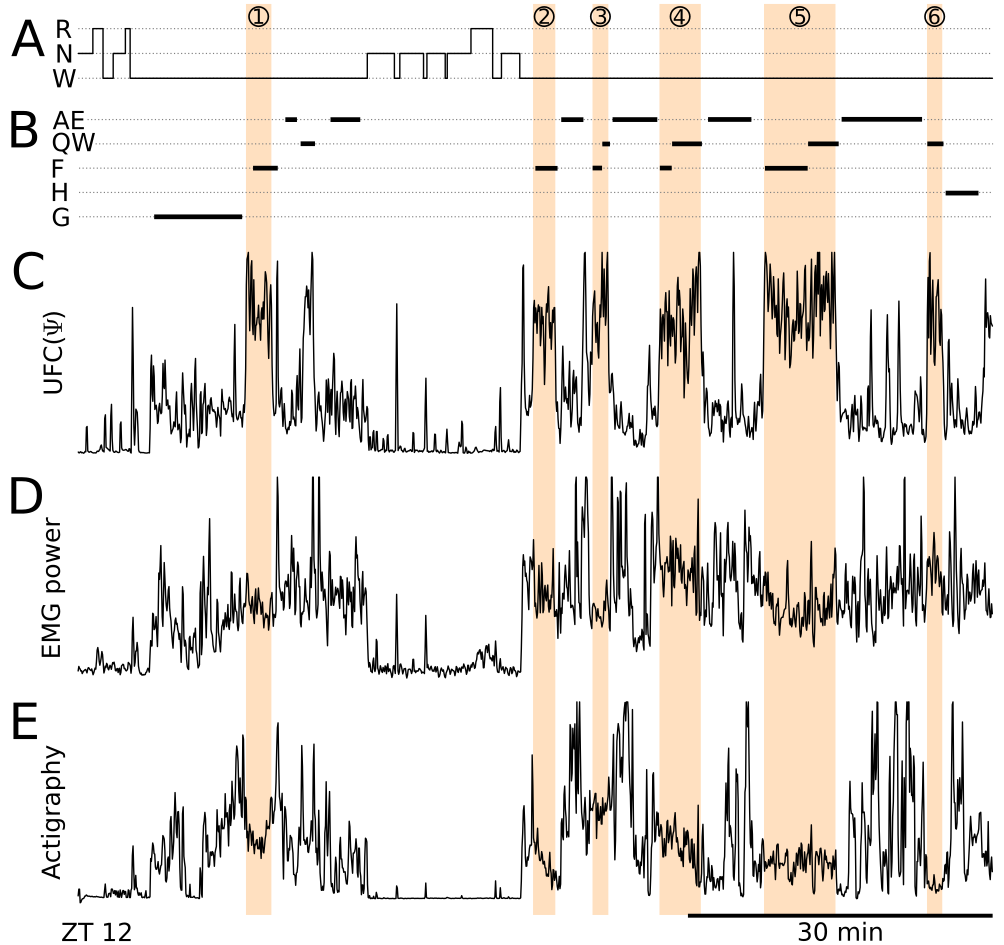}
  \caption{\linespread{1.2} \footnotesize\textbf{(Supplementary) Ultra fast component of $\boldsymbol{\Psi_{EEG}}$ and its behavioral correlates during wakefulness.} The activity patterns of the first 90 minutes after lights-off (ZT 12) are shown for a representative animal (M12, table \ref{tab:animals} in methods). \textbf{A.} Hypnogram scored based on EEG/EMG (R = REM sleep; N = NREM sleep; W = wakefulness). \textbf{B.} Annotations of behavioral states within wakefulness. AE = active exploration (fast locomotion around the cage, commonly standing up against the cage wall]). QW = quiet-wake (slower movements, frequent still vigilant position, sniffing with the nose pointing up). F = feeding. H = housekeeping (rearranging the bedding material). G = grooming. Behavior records are discontinuous as only unambiguous periods of a stereotyped behavior were tagged (as judged from a time-lapse sequence of images; see methods). \textbf{C.} Ultra fast component of $\Psi_{\mathit{EEG}}$ (UFC($\Psi$)). Six colored boxes (\textcircled{1}-\textcircled{6}) highlight periods when UFC($\Psi$) was over its quantile 0.75 for more than 1.5 minutes. \textbf{D.} EMG power (70-90 Hz). UFC($\Psi$) is plotted in linear scale to appreciate the pronounced fluctuations of the variable during wakefulness (compare with fig. \ref{fig:UFC}). \textbf{E.} Actigraphy based on time-lapse photography (see methods).  Notice that in high UFC($\Psi$) the animal is mostly engaged in quiet-wake and feeding behaviors. The fact that quiet-wake shares the higher quantiles of UFC($\Psi$) with feeding behavior helps to discard the hypothesis that high  UFC($\Psi$) could be related to the activity of the masticatory musculature. It is common that high UFC($\Psi$) periods start with feeding behavior and end with quiet-wake (e.g. \textcircled{3}, \textcircled{4}, and \textcircled{5}).}   
  \label{fig:actigraphy}
\end{figure}

\newpage
%---------------------------------------------------------------------------------
\begin{figure}[H]
  \centering
    \includegraphics[width=8.5cm, keepaspectratio]{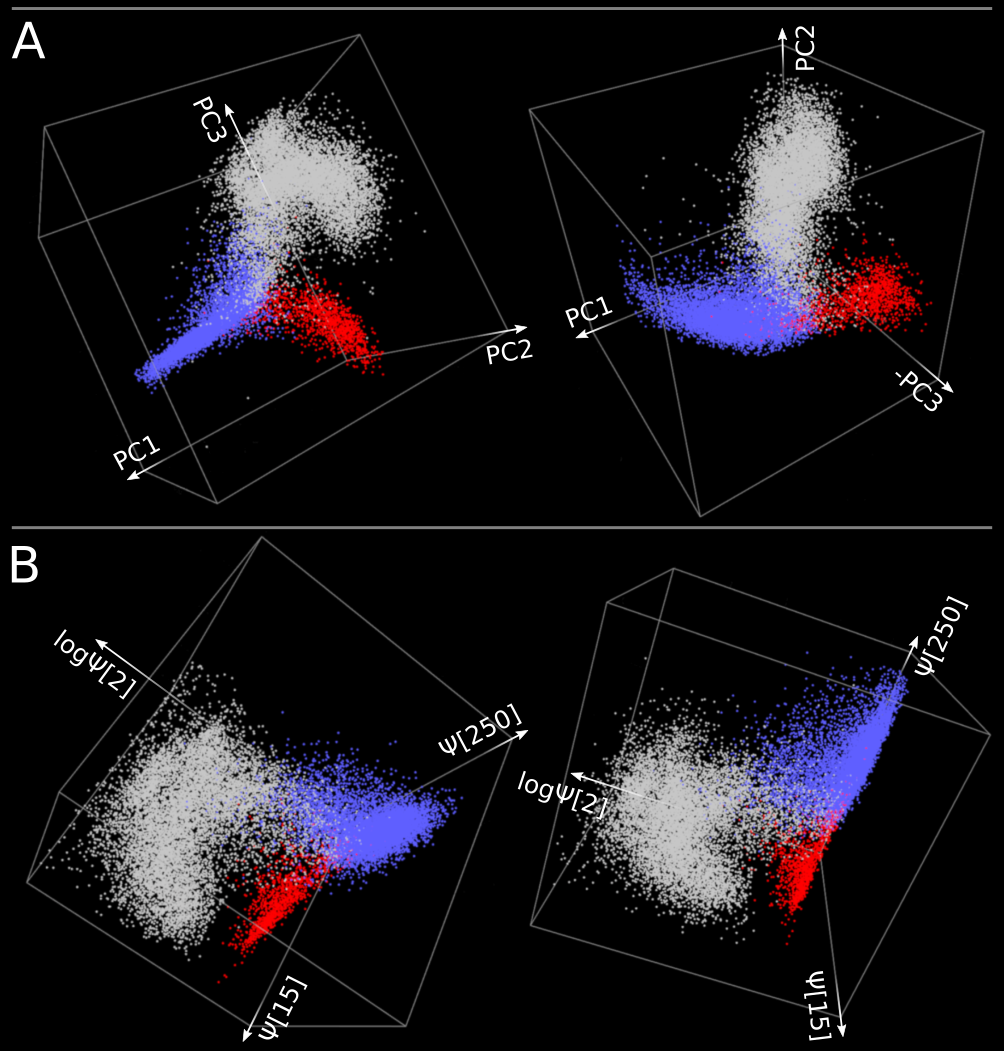}
  \caption{\linespread{1.2} \footnotesize\textbf{(Supplementary) Clustering of behavioral states using raw coordinates of $\boldsymbol{\Psi_{EEG}}$-space.} \textbf{A.} PCA of $\Psi_{EEG}$-space calculated in the same conditions as indicated in figure \ref{fig:PCA}B using a different representative animal (M16, table \ref{tab:animals} in methods) gives an analogous result. \textbf{B.} From the same $\Psi_{EEG}$-space, a 3D subspace is built using only the raw $\Psi-$ pattern values at samples  $2$, $15$, and $250$. These three coordinates represent respectively the ultra-fast spike related to wakefulness (gray), the fast component related to REM sleep (red) and wakefulness, and the slow component related to NREM sleep (blue).} \label{fig:PCA2} 
\end{figure}

%---------------------------------------------
\newpage
\section*{Methods}
\subsection*{Animals}
All experimental procedures were carried out in accordance with local and national regulations after approval by the University of Tsukuba's animal care and use committee. Our experiments use male C57BL/6J mice (The Jackson Laboratory, Japan) aged $10$-$25$ weeks. The current manuscript's findings are based on the experimental data summarized in table \ref{tab:animals}. This data is a subset of a larger set of experiments in which only high-quality EEG/EMG recordings were chosen, with special emphasis on the rejection of 50 Hz powerline interference without the use of a notch filter (see below for details).

\begin{table}[!htb]
\centering
\caption{Experimental animals.}
\label{tab:animals}
\small
\begin{tabular}{llSS}
\hline
\multicolumn{1}{c}{\textbf{Mouse ID}} & \textbf{Experiment type} & \textbf{recording time (continuous days)} & \textbf{Sampling rate (kHz)}\\ \hline
M1 & Spontaneous EEG & 6 & 5.0 \\
M2 & Spontaneous EEG & 6 & 5.0 \\
M3 & Spontaneous EEG & 7 & 5.0 \\
M4 & Spontaneous EEG & 7 & 5.0 \\
M5 & Spontaneous EEG & 6 & 5.0 \\
M6 & Spontaneous EEG & 2 & 5.0 \\
M7 & Spontaneous EEG & 1 & 5.0 \\
M8 & Spontaneous EEG & 3 & 5.0 \\
M9 & Spontaneous EEG & 3 & 5.0 \\
M10 & Spontaneous EEG & 10 & 5.0 \\
M11 & Spontaneous EEG & 7 & 5.0 \\
M12 & Spontaneous EEG & 14 & 5.0 \\
M13 & Spontaneous EEG & 14 & 5.0 \\
M14 & Spontaneous EEG & 4 & 5.0 \\
M15 & Spontaneous EEG & 3 & 5.0 \\
M16 & Spontaneous EEG & 1 & 12.5 \\
M17 & Spontaneous EEG & 2 & 12.5 \\
M18 & Spontaneous EEG (SD baseline) & 4 & 5.0 \\
M18 & Sleep deprivation & 2 & 5.0 \\
M19 & Spontaneous EEG (SD baseline) & 4 & 5.0 \\
M19 & Sleep deprivation & 2 & 5.0 \\ \hline
\end{tabular}
\end{table}

\subsection*{EEG surgery}
After inducing anesthesia with isoflurane $3.5\%$ for $2$ min, mice were placed on a stereotaxic frame (David Kopf Instruments, Tujunga, CA, USA) under isoflurane $2\%$. In this condition, custom-made EEG and EMG electrodes were implanted aseptically. EEG electrode screws were placed epidurally on the right frontal and parietal cortices (AP, $+1$ mm to the front of bregma or lambda; LR $-1.5$ mm lateral from the sagittal raphe) according to the mouse brain atlas by \citet{franklin2019paxinos}. Two EMG Teflon-coated electrodes were inserted bilaterally into the nuchal muscles. Finally, the electrode assembly attached to a 4-pin connector was fixed to the skull with self-curing dental cement. Sutures were then applied to the incised skin. After surgery, the animals were administered an intraperitoneal injection of ampicillin (100 mg/kg) and allowed to recover in individual housing cages for one week. Each animal was then attached to an EEG/EMG recording cable, for acclimatization to an individual soundproof chamber for an additional three days. The cables are long and flexible enough to minimize movement restrictions for the mice within their cages.

\subsection*{EEG/EMG acquisition} EEG and EMG were acquired using INTAN amplifier RHD2132 (16-channel headstage PN3334; INTAN Technologies, LA, USA) usually at 5kHz (low-pass filter 2.5 kHz, high-pass filter 0.1 Hz) and for special experiments (see fig. \ref{fig:UFC}J) at 12.5 kHz (low-pass filter 6 KHz, high-pass filter 0.1 Hz) and stored as RHD files (1 file per hour of recording). RHD files were loaded using custom R functions for offline analysis (R Foundation for Statistical Computing, Vienna, Austria; \textit{http://www.R-project.org/}). These functions were developed to extend the functionality of the python code provided by INTAN for reading RHD files, which was integrated into the R workflow via the package \textit{reticulate}. To avoid $50$ Hz powerline interference, an aluminum tube ($35$ mm length, $16$ mm external diameter, and $14$ mm internal diameter) was used to shield the INTAN headstage (grounded to the headstage's GND pin). Analog and digital notch filters at $50$ Hz were avoided because they distort the structure of $\Psi_{EEG}$ patterns.

\subsection*{EEG staging}
Using a custom Matlab-based sleep scoring software, a human expert visually classified the raw EEG data segmented into $10$-second epochs into Wake, NREM, or REM states using well-established conventional rules for rodent EEG \citep{Robert1999-wg}. The special label "Artifact" was assigned to epochs that had motion artifacts or any other EEG artifact that obscured the archetypal EEG signatures of physiological brain states. Using a simple function, 10-s scoring vectors were mapped to 4-s scoring vectors.

{\footnotesize
\begin{lstlisting}[xleftmargin = 1.5em]
map.vector <- function(v, new.length){
  l <- length(v)
  v[ceiling(1:new.length*l/new.length)]
}

score.4s.24h <- map.vector(score.10s.24h, 21600)
\end{lstlisting}
}

\subsection*{Sleep deprivation}
While an EEG was being recorded, mice were sleep-deprived (SD) for $6$ hours using a gentle handling procedure that began at the light's onset (Zeitgeber time (ZT) = $0$, at 9:00 AM) and ended at ZT = $6$ (3:00 PM). Gentle handling techniques include gently disturbing the bedding materials, tapping the mouse cage, and gently poking the animal with soft materials when the animal behaves sleepily. Mice were allowed to enter sleep recovery for 42 hours (completing 2 days from the beginning of the protocol). Baseline recordings were collected for 96 hours prior to the SD intervention.

\subsection*{Time-lapse photography based actigraphy}
An infrared-sensitive camera (Raspberry Pi NoIR Camera V2) was attached to the CSI port on a Raspberry Pi 3 Model B+  equipped with a real-time clock (based on Maxim DS3231 RTC IC) to allow synchronization with the main computer attached to the INTAN acquisition board. Using commands provided by the Raspberry Pi camera libraries (i.e. \textit{raspistill}) time-lapse pictures (640x480 pixels) were taken continuously every second and stored in the Raspberry Pi SD card. The starting time of the sequence was automatically scheduled using the Linux utility \textit{cron}. Every picture file was postmarked with a string composed of the date/time info allowing the stability of the frame stream to be tracked (no noticeable deviations over periods of 24 hours at 1-second resolution). The time-lapse sequence was used to visually score the behavior of the animal during wakefulness (e.g. active exploration, quiet-wake, feeding, housekeeping, and grooming) and to generate an actigraphy record by fusing  the RGB channels of each frame into a single monochromatic channel by simple averaging (i.e. 640x480x3 array to 640x480 matrix). In order to assess the activity patterns of the animal, each matrix (monochrome frame) was subtracted from the previous matrix in the sequence, and the result was integrated (i.e. sum(current\_frame - previous\_frame)). The time-lapse sequence produces 8 values per epoch (4-second epochs with 50\% overlap), which were averaged to create an actigraphic value per epoch.

\subsection*{Pulse superposition (signal generation)}
A numerical implementation in R of the stochastic superposition $X$ (eq. \ref{eq:model}) is produced by the circular convolution of a vector representing the pulse $g(t)$ with a second vector representing a Poisson Process ($\mathit{PP}$).

{\footnotesize
\begin{lstlisting}[xleftmargin = 1.5em]
X <- convolve(pulse, rev(PP), type = "circular")
\end{lstlisting}
}

\noindent
To create a $PP$, a simple method is to generate a vector of random values from a Poisson distribution controlled by parameter $\lambda$ which represents the events per second. At the sample level, the expected rate of event occurrences in the Poisson distribution is given by $\lambda/f_s$, where $f_s$ is the sampling rate.

{\footnotesize
\begin{lstlisting}[xleftmargin = 1.5em]
PP <- rpois(n.samples, lambda/fs)
\end{lstlisting}
}

\noindent
When $\lambda$ is a constant, the Poisson process is homogeneous. Alternatively, a non-homogeneous Poisson process can be generated using a vector (same length as $PP$) representing temporal $\lambda$ variations along $PP$ (see fig. \ref{fig:nonHomogPoisson}).

Alternatively, $PP$ can be generated by randomly positioning each event in time following a uniform distribution ---conceptually an event here is a discrete impulse function. This algorithm is slower but allows the amplitudes to be scaled arbitrarily for each impulse. For example, in the function \textbf{gen.shot.noise} the vector \textbf{amp.vect} represents the amplitude of every event composing the $PP$. In the following code example, $PP$ gets a length of \textbf{n.samples} and an amount of \textbf{n.pulses} (i.e. the length of \textbf{amp.vect}) which are positioned among those samples, each with a random amplitude taken from an exponential distribution (see supplementary fig. \ref{fig:scaledImpulses}).

{\footnotesize
\begin{lstlisting}[xleftmargin = 1.5em]
gen.shot.noise <- function(n.samples, amp.vect){
  n.pulses <- length(amp.vect)
  p.arrival <- sample(1:n.samples, n.pulses, replace = T)
  superpos <- numeric(n.samples)
  for(i in 1:n.pulses){
    superpos[p.arrival[i]] <- superpos[p.arrival[i]] + amp.vect[i]
  }
  superpos
}

PP <- gen.shot.noise(n.samples, amp.vect = rexp(n.pulses))
\end{lstlisting}
}

Obtaining $PP$ via \textbf{gen.shot.noise} can reproduce the result of generating $PP$ via \textbf{rpois} assigning a unitary amplitude to every event.

{\footnotesize
\begin{lstlisting}[xleftmargin = 1.5em]
PP <- gen.shot.noise(n.samples, amp.vect = rep(1, n.pulses))
\end{lstlisting}
}

Finally, as $X$ is used as a model of an AC signal (e.g. EEG), any DC component is discarded (i.e. $X = X - mean(X)$).

\subsection*{Special pulse functions}
The following R code demonstrates the implementation of alpha functions and dual-exponential functions in detail. The parameter \textbf{tau} is the temporal constant $\tau$, \textbf{fs} is the sampling rate, and \textbf{t.max} is the desired time length in seconds. In the dual-exponential function the parameters \textbf{tau1} and \textbf{tau2} control the dynamics of the rising phase and falling phase respectively and the condition \textbf{tau1} $<$ \textbf{tau2} must be satisfied.

{\footnotesize
\begin{lstlisting}[xleftmargin = 1.5em]
do.alpha.pulse <- function(tau, fs, t.max){
  time <- seq(from=0, by=1/fs, length.out = fs*t.max)
  pmax(time/tau*exp(1-time/tau), 0)
}

do.dual.exp.pulse <- function(tau1, tau2, fs, t.max){
  time <- seq(from=0, by=1/fs, length.out = fs*t.max)
  pmax(tau1*tau2/(tau1 + tau2)*(exp(-time/tau2) - exp(-time/tau1)), 0)
}
\end{lstlisting}
}

For simplicity related to signal power considerations, it is convenient to generate pulses with unit energy ($E_g = 1$) or otherwise rescale them to have unit energy. The function \textbf{unitE} rescales the input vector to a unit energy vector by default or to any other energy value determined by the parameter \textbf{outputE}.

{\footnotesize
\begin{lstlisting}[xleftmargin = 1.5em]
unitE <- function(v, outputE = 1){
  inputE <- sum(v^2)
  v*sqrt(outputE/inputE)
}
\end{lstlisting}
}

For example, the following code generates a vector representing an alpha function $\tau = 0.05$, $f_s = 1000$ samples per second, and $200$ ms in length. After its creation, the vector is constrained to have unit energy.

{\footnotesize
\begin{lstlisting}[xleftmargin = 1.5em]
pulse <- do.alpha.pulse(0.05, 1000, 0.2)
pulse <- unitE(pulse)
\end{lstlisting}
}

Making $E_g = 1$, the power of the superposition $X$ is controlled only by $\lambda_g$. In addition, generating $X$ using a different kind of pulse at the same $\lambda_g$, allows comparing  iso-energetic signals where any difference in $\Psi$ analysis (eq. \ref{eq:PsiDef}) only depends on the shape of the generating pulse.

\subsection*{Superposition of pulse mixtures}
For simplicity of algorithms, the generation of superpositions of mixtures $X_{\mathit{mix}}$ due to generating pulses $g_1(t),$ $g_2(t), ..., g_n(t)$ was achieved by creating superpositions of same-class pulses first --- i.e. $X_i(t) = \sum^N_{j=1} g_i(t - t_j)$ --- and then adding them all together --- i.e. $X_{\mathit{mix}} = X_1 + X_2 + ... + X_n$. Therefore, the energy contribution of each type of pulse $g_n(t)$ is simply the energy fraction of $X_n$ relative to the overall energy of $X_{\mathit{mix}}$, which is determined by the product $[\lambda_g E_g]_i$.

\subsection*{Calculation of $\Psi_{EEG}$}
The following code is a minimal implementation of $\Psi_{EEG}(t^{\prime}) =  \frac{1}{2}\frac{d}{dt^{\prime}}\sigma^2_{\mathit{\widehat{EEG}}}(t^{\prime})$ (eq. \ref{eq:PsiDef}). In the code, \textbf{epoch} corresponds to a particular EEG epoch, and \textbf{max.delay} corresponds to the maximum desired value for $t^{\prime}$.

{\footnotesize
\begin{lstlisting}[xleftmargin = 1.5em]
var.delayed <- function(s, max.delay){
  var.d <- numeric(max.delay)
  for(i in 1:max.delay) var.d[i] <- var(diff(s,i))
  var.d
}

Psi_epoch <- diff(var.delayed(epoch,max.delay))/2
\end{lstlisting}
}

Alternatively, $\Psi_{EEG}$ (per epoch) can be calculated via the autocovariance function (eq. \ref{eq:PsiDef}). Using the built-in \textbf{acf} function in R, an implementation of $\Psi_{EEG}(t^{\prime}) = -\frac{d}{dt^{\prime}} \gamma_{EEG}(t^{\prime})$ is indicated in the following code.

{\footnotesize
\begin{lstlisting}[xleftmargin = 1.5em]
neg.diff.ACF <- function(v, max.delay){
   -diff(c(acf(v, max.delay, type= "covariance", plot=F)$acf))
}

Psi_epoch <-  neg.diff.ACF(epoch, max.delay)
\end{lstlisting}
}

\subsection*{$\Psi_{EEG}$ dynamics and its colored map representation.}
To analyze dynamic changes in non-stationary signals (e.g. EEG), the signal must be divided into epochs. The general purpose function \textbf{epoch.feature} divides the input vector \textbf{v} into epochs of length \textbf{epoch.} \textbf{length} (seconds), with overlap defined by \textbf{epoch.overlap} (seconds), and \textbf{fs} is the sampling rate. An \textit{epoch feature} characterizing each epoch --- e.g. the $\Psi$-pattern ---  is extracted by the function \textbf{func}. If \textbf{func} returns a scalar (e.g. by calculating the energy per epoch), \textbf{epoch.feature} returns a vector of length equal to the number of epochs $N$. On the other hand, when \textbf{func} returns a vector of length $L$ (e.g. by calculating $\Psi$ per epoch), \textbf{epoch.feature} returns a matrix of dimension $N \times L$. The extension of epochs required by the overlap at both ends of the input vector is solved by the function \textbf{specular.ext}. A segmentation example of 4-second epochs with 50\% overlap is shown in the last line of the following code, generating a $\Psi_{EEG}$-matrix with dimensions $21400 \times \mathit{max.delay}$

{\footnotesize
\begin{lstlisting}[xleftmargin = 1.5em]
specular.ext <- function(K, lim.sup){
  N <- length(K)
  L <- sum(K < 1)
  if(L != 0) K[1:L] <- (L + 1):2
  L <- sum(K > lim.sup)
  if(L != 0) K[(N - L + 1):N] <- (lim.sup - 1):(lim.sup - L)
  K
}

epoch.feature <- function(v, epoch.length, epoch.overlap, fs, func){
  n <- trunc(length(v)/fs/epoch.length)
  feature <- list()
  rng.1 <- (-epoch.overlap*fs + 1):(epoch.length*fs + epoch.overlap*fs)
  pb <- txtProgressBar(style=3,width=20)
  for(i in 1:n){
    rng <- specular.ext(rng.1 + (i - 1)*epoch.length*fs, length(v))
    sub.v <- v[rng]
    feature[[i]] <- func(sub.v)
    setTxtProgressBar(pb, i/n)
  }
  close(pb)
  len.feat <- length(feature[[1]])
  feature <- unlist(feature)
  if(len.feat > 1) feature <- matrix(feature, nrow=n, byrow=T)
  feature
}

Psi_EEG <- epoch.feature(EEG, 4, 2, 5000,
            function(epoch) neg.diff.ACF(epoch, max.delay))
\end{lstlisting}
}

The matrix $\Psi_{EEG}$, which contains the $\Psi$-pattern characterizing each EEG epoch, can be directly visualized as a colored map (as customary for spectrograms). Given that \textit{zero} is meaningful in $\Psi$ scale, it is convenient to map the value \textit{zero} to a particular pure color. For example, in our pseudo-color scheme, the value \textit{zero} is mapped to black. A gradient [black $\rightarrow$ blue $\rightarrow$ cyan] indicates increasing positive values, while a gradient [black $\rightarrow$ dark red] indicates decreasing negative values (fig. \ref{fig:EEG_analysis}E).

\subsection*{Ultra fast and slow $\Psi$ components}
 When sampling EEG at $5$ kHz, we define the ultra-fast component of $\Psi$ (UFC($\Psi$)) as the integration of the first 3 elements of the vector $\Psi$. Similarly, the slow component of $\Psi$ (SC($\Psi$)) is defined by the integration of energy during the first 500 samples of $\Psi$ --- at $5$ kHz which corresponds to the first $100$ ms. It is worth noticing that according to eq. \ref{eq:PsiDef}, UFC($\Psi$) and SC($\Psi$) can be directly obtained by $\sigma^2(\mathit{epoch}(t) - \mathit{epoch}(t + t^{\prime}))/2$, with $t^{\prime} = 3$ samples and $t^{\prime} = 500$ samples, respectively.

\subsection*{Multitaper spectrogram}
Multitaper spectrogram theory and applications to EEG are well explained in \citet{Prerau2017-ov}, and their methodology is used here. Tapers (Slepian sequences) were generated in R using the package \textit{multitaper}. Before applying Fourier analysis EEG was down-sampled from 5kHz to 500Hz using the \textit{resample} function from the package \textit{signal}. The spectrogram is evaluated in logarithmic scale for visualization purposes and in linear scale for spectral band power estimation.

\subsection*{Spectral band power}
Spectral power, a uni-variate feature per epoch, is calculated by integrating the power associated with a particular spectral band, i.e. the power spectrum area within given band limits. EEG delta band was evaluated between $0.5 - 4$ Hz. EEG theta band was evaluated between $6 - 12$ Hz, and EMG power was evaluated between $70 - 90$ Hz.

\subsection*{Principal component analysis}
Principal component analysis (PCA) of the $\Psi_{EEG}$ was performed using the first 15 ms ($t^{\prime}$) per epoch (which corresponds, working at 5kHz, to evaluate $\Psi_{EEG}$ using \textit{max.delay} = $75$ samples). Thus, the $\Psi_{EEG}$ matrix (24h EEG) has a dimensionality of $21400$ x $75$. The function \textit{prcomp} (R language) was applied over $\Psi_{EEG}$ using standard options (i.e. variables were zero-centered and scaled to unit variance). As illustrated in figure \ref{fig:PCA}, the projection (dot product) of $\Psi_{EEG}$ over each of the first $3$ principal components provides the 3D-subspace shown (3D plots generated using the package \textit{rgl} in R).
%---------------------------------------------
\newpage
\begin{singlespace}
\bibliographystyle{apalike}
\bibliography{References}
\end{singlespace}
%---------------------------------------------
\newpage
\section*{Supplementary material}
\subsection*{Basic relationship between the Variance and the Power of a signal.}
Considering a signal $X$ with $N$ samples ${x_1, x_2, x_3, ..., x_N}$, the variance (population variance) of such signal is
\begin{equation}
	\sigma^2_X = \frac{1}{N} \sum_{i=1}^{N} (x_i - \overline{X})^2
\end{equation}
\noindent
for a signal centered in zero ($\overline{X} = 0$) the expression simplifies to
\begin{equation}
	\sigma^2_X = \frac{1}{N} \sum_{i=1}^{N} x_i^2
\end{equation}
The energy of $X$ has a similar definition
\begin{equation}
	E_X = \sum_{i=1}^{N} x_i^2
\end{equation}
\noindent
the variance and energy of $X$, therefore, simply correspond as
\begin{equation}
	\sigma^2_X = \frac{E_X}{N}
\end{equation}
The power of $X$ is ($t$ is time)
\begin{equation}
	P_X = \frac{E_X}{t}
\end{equation}
\noindent
Then, the relationship between power and variance can be described as
\begin{equation}
	\sigma^2_X = \frac{P_X \cdot t}{N}
\end{equation}
\noindent
and because the sampling rate $f_s$ is determined by the ratio $N/t$, the expression becomes
\begin{equation} \label{eq:sup_varPow}
	\sigma^2_X = \frac{P_X}{f_s}
\end{equation}

%-------------------------------------------------------
\subsection*{Important property of Poisson distribution}

The Poisson distribution probability mass function ($k \in \mathbb{N}\:; \lambda$ indicates event density) is
\begin{equation}
f(k, \lambda) = \frac{\lambda^k e^{-\lambda}}{k!}
\end{equation}

Let $y$ be a set of $N$ deviates following a Poisson distribution with parameter $\lambda$. An important property of the Poisson distribution is that both the mean and the variance are equal to $\lambda$ \citep[p.~64]{Linden2013-qb}.
\begin{equation}
\bar{y} = \sigma^2_y = \lambda
\end{equation}

Regarding our model, the average can be directly interpreted as event density (e.g. events per second) and the variance as the signal power, according to eq. \ref{eq:sup_varPow}. Interestingly, this implies that the power of the AC component produced in a Poisson process is simply the amount of energy carried by the count of events per second. For example, in our model $P_X = \mathit{E_g} \lambda_g$, $P_X$ is the power of process $X$, $E_g$ is the energy of the pulse $g(t)$, and $\lambda_g$ is the pulse density. Despite what might seem to be an obvious assertion regarding energy conservation, pulse trains following non-Poisson distributions might not satisfy this condition. For instance, a \textit{super-Poisson distribution} (i.e. $\sigma^2_y > \bar{y}$) with the same rate of pulses is expected to have a larger power due to more frequent occurrences of high-pulse-density bins given that these higher figures are squared when calculating power.

%-------------------------------------------------------
\subsection*{Variance of $\boldsymbol{\hat{X}}$}

The covariance between discrete processes $X$ and $Y$ is
\begin{equation}
\operatorname{Cov}(X,Y) = \frac{1}{N} \sum_{i=1}^N x_i \; y_i - \Big( \sum_{i=1}^N x_i \Big) \Big( \sum_{i=1}^N y_i \Big)
\end{equation}

\noindent
now considering the autocovariance $\gamma$ of a process $X$ centered in 0 (for example, an AC signal), simply reduce to ($k$ is the delay)

\begin{equation}
\gamma_X(k) = \frac{1}{N} \sum_{i=1}^{N-k} x_i \; x_{i+k}
\end{equation}

The variance of $\hat{X}$ is expressed as
\begin{equation}
\sigma_{\hat X}^2(k) = \frac{1}{N} \sum_{i=1}^{N-k} (x_i - x_{i+k})^2
\end{equation}

\begin{equation}
\sigma_{\hat X}^2(k) = \frac{1}{N} \sum_{i=1}^{N-k} x^2_i -2 x_i \; x_{i+k} + x^2_{i+k}
\end{equation}

if $N \gg k$ then

\begin{equation}
\sum_{i=1}^{N-k} x^2_i \simeq \sum_{i=1}^{N-k} x^2_{i+k}
\end{equation}

\begin{equation}
\sigma_{\hat X}^2(k) \simeq \frac{2}{N} \sum_{i=1}^{N-k} x^2_i - \frac{2}{N} \sum_{i=1}^{N-k} x_i \; x_{i+k}
\end{equation}

\begin{equation}
\sigma_{\hat X}^2(k) \simeq 2(\sigma_X^2 - \gamma_X(k))
\end{equation}

\begin{equation}
\lim_{N\to\infty}\; \sigma_{\hat X}^2(k) = 2(\sigma_X^2 - \gamma_X(k))
\end{equation}

In addition, the variance of $\hat X$ as a function of $g(t)$ can be calculated as follows. Let us begin with the definition of $X$.

\begin{equation} \label{eq:sup_model}
X(t) = \sum^N_{j=1} g(t - t_j)
\end{equation}

\noindent
Given that $X$ is a Poisson process with a pulse density $\lambda_g$ (i.e. events per second), the power of $X$ is

\begin{equation}
P_X = \mathit{E_g} \lambda_g
\end{equation}

According to eq. \ref{eq:sup_varPow}, this power can be related to the variance of process $X$.

\begin{equation} \label{eq:sup_var&Pow}
\sigma_X^2 = \frac{P_X}{fs} = \frac{\lambda_g}{f_s} E_g
\end{equation}

The analysis of the power of $X$ can be improved by studying $\hat{X}(t^{\prime}) = X(t) - X(t + t^{\prime})$. Given the definition of $X$, $\hat{X}$ can be understood as the stochastic superposition of the  pulse $g(t - t_j) - g(t - t_j + t^{\prime})$ obtained by combining $g(t)$ with it's delayed version.
\begin{eqnarray} \label{eq:sup_Xhat}
        \hat{X}(t^{\prime}) &=& X(t) - X(t + t^{\prime}) \\
        &=& \sum^N_{j=1} g(t - t_j) - \sum^N_{j=1} g(t - t_j + t^{\prime}) \\
        &=& \sum^N_{j=1} g(t - t_j) - g(t - t_j + t^{\prime})
\end{eqnarray}
\noindent
The energy of $g(t - t_j) - g(t - t_j + t^{\prime})$, which is  independent of time  location $t_j$, fulfills the following relation
\begin{eqnarray} \label{eq:sup_Xhat2}
        \int_{-\infty}^\infty \Big( g(t) - g(t + t^{\prime}) \Big)^2 dt &=& 2 \int_{-\infty}^\infty g^2(t) dt - 2 \int_{-\infty}^\infty g(t)g(t + t^{\prime}) dt \\ &=& 2 \Big( E_g - (g \star g)(t^{\prime}) \Big)
\end{eqnarray}
(note: $\star$ denotes the autocovariance not the autocorrelation)
\noindent
considering the pulse density $\lambda_g$, the power of $\hat{X}$ as a function of $t^{\prime}$ is
\begin{equation}
P_{\hat{X}}(t^{\prime}) = 2 \lambda_g \Big( E_g - (g \star g)(t^{\prime}) \Big)
\end{equation}

combining with eq. \ref{eq:sup_varPow}, the variance of $\hat{X}$ as a function of $t^{\prime}$ can be stated as 

\begin{equation} \label{eq:sup_varXtongo}
\sigma_{\hat{X}}^2(t^{\prime}) = 2\frac{\lambda_g}{f_s} \Big( E_g - (g \star g)(t^{\prime}) \Big)
\end{equation}

Finally, combining this expression with equations \ref{eq:var&Pow} and \ref{eq:autoCov1}, the autocovariance of $X$ can be expressed as

\begin{equation}
\gamma_X(t^\prime) = \frac{\lambda_g}{f_s} \Big((g \star g)(t^{\prime}) \Big)
\end{equation}

These results are analogous to Campbell's seminal findings on the analysis of the continuous shot noise process, whose rigorous derivation may be found in \citet[p.~126]{Linden2013-qb} and \citet{campbell1909}. 

%(poner referencias originales de campbell 1909, 1910 y paper de 1971).

%-------------------------------------------------------------------
\subsection*{Mathematical proofs regarding special pulses in fig. \ref{fig:pulses} upper row.}
Figure \ref{fig:pulses}'s top row depicts four pulses $g(t)$ --- namely exponential decay, alpha function, dual-exponential function, and square pulse --- whose exact shape can be recovered using the $\Psi$ operator. This is possible because these four functions fulfill the following admissibility condition linking the integral of pulses with their autocorrelation function  ($k$ is an scaling factor).

\begin{equation}
E_g - (g \star g)(t^\prime) = k \int_0^{t^\prime} g(t)dt
\end{equation}

which is equivalent to

\begin{equation} \label{eq:sup_toProof}
\int_0^\infty g^2(t) dt - \int_0^\infty g(t)g(t + t^{\prime}) dt = k \int_0^{t^\prime} g(t)dt
\end{equation}

below we provide the proofs how these four pulses fulfill the condition. 

\subsubsection*{Proof 1. Exponential decay.}
The LHS of eq. \ref{eq:sup_toProof} denoted by $A$ is described as follows when $g(t)=\exp(-\alpha t)$.
\begin{eqnarray}
        A &=&\int^\infty_0 \exp(-2\alpha t) dt - \int^\infty_0 \exp(-\alpha t) \exp(-\alpha (t + t^{\prime}))dt \\
        &=& (1- \exp(-\alpha t^{\prime}))\int^\infty_0 \exp(-2\alpha t) dt \\
        &=& \frac{1}{2\alpha} (1- \exp(-\alpha t^{\prime}))
\end{eqnarray}

Moreover, the RHS of eq. (10) denoted by $B$ is as follows.
\begin{eqnarray}
        B&=&k\int^{t^{\prime}}_0 \exp (-\alpha t) dt \\ &=&\frac{k}{\alpha} (1-\exp(-\alpha t^{\prime}))
\end{eqnarray}
Therefore, $A=B$ implies that $\displaystyle k=\frac{1}{2}$.

\subsubsection*{Proof 2. Alpha function.}
The LHS of eq. \ref{eq:sup_toProof} denoted by $A$ is described as follows when $g(t) = t\exp(-t)$.
\begin{eqnarray}
        A &=&\int^\infty_0 t^2\exp(-2 t) dt - \int^\infty_0 t (t +t^{\prime} ) \exp(- t) \exp(- (t+t^{\prime}))dt \\
        &=& (1-\exp (-t^{\prime}))\int^\infty_0 t^2 \exp(-2t) - t^{\prime} \exp(-t^{\prime})\int^\infty_0 t \exp (-2t)dt \\
        &=&(1-\exp (-t^{\prime})- t^{\prime}\exp (-t^{\prime}))\int^\infty_0 t \exp (-2t)dt \\ 
        &=& \frac{1}{4}(1-\exp (-t^{\prime})- t^{\prime}\exp(-t^{\prime}))
\end{eqnarray}

Moreover, the RHS of eq. (10) denoted by $B$ is as follows.
\begin{eqnarray}
        B&=&k\int^{t^{\prime}}_0 t \exp (-t) dt \\ 
        &=& k [-t^{\prime} \exp (-t^{\prime})+ 1 - \exp(-t^{\prime})]
\end{eqnarray}

Therefore, $A=B$ implies that $\displaystyle k=\frac{1}{4}$.
\subsubsection*{Proof 3. Dual-exponential pulse.}
The LHS of eq. \ref{eq:sup_toProof} denoted by $A$ is described as follows when $g(t)=C(\exp(-\alpha t)- \exp (-\beta t))$.
Here $C=\frac{1}{\beta-\alpha}$.
\begin{eqnarray}
        A &=&\int^\infty_0 C^2 (\exp(-\alpha t)- \exp (-\beta t))^2 dt  \nonumber \\ & &- \int^\infty_0 C^2(\exp(-\alpha t)- \exp (-\beta t))\times(\exp(-\alpha (t+t'))- \exp (-\beta (t+t'))) dt, 
\end{eqnarray}
where the first term is denoted by $A_1$ and the second is $A_2$, namely $A=A_1-A_2$. In addition, we denote $\Gamma_1 = \exp(-\alpha t')$ and $\Gamma_2 = \exp(-\beta t')$. 

\begin{eqnarray}
        A_1 &=& C^2 \int^\infty_0 [\exp(-2\alpha t)+ \exp(-2\beta t) -2 \exp(-(\alpha+\beta)t)] dt \nonumber \\
        &=& C^2 \left[-\frac{1}{2\alpha}\exp(-2\alpha t) -\frac{1}{2\beta}\exp(-2\beta t) +\frac{2}{\alpha+\beta}\exp(-(\alpha +\beta)t)\right]^\infty_0 \nonumber \\
        &=& C^2\left( \frac{1}{2\alpha}+ \frac{1}{2\beta} - \frac{2}{\alpha+\beta}\right)\nonumber \\
        &=& \frac{1}{(\beta-\alpha)^2}\left(\frac{\alpha+\beta}{2\alpha\beta}-\frac{2}{\alpha+\beta} \right) \nonumber\\
        &=&\frac{1}{2\alpha\beta(\alpha+\beta)} \nonumber
\end{eqnarray}

Next, we calculate $A_2$ as follows.
\begin{eqnarray}
        A_2 &=& C^2 \int^\infty_0 \left\{\exp(-\alpha t)+ \exp(-\beta t)\right\}\left\{ \Gamma_1\exp(-\alpha t)-\Gamma_2 \exp(-\beta t)\right\} dt \nonumber \\
        &=&C^2 \int^\infty_0 \left[\Gamma_1\exp(-2\alpha t) + \Gamma_2 \exp(-2 \beta t) -(\Gamma_1+\Gamma_2)\exp(-(\alpha+\beta)t)\right] dt \nonumber\\
        &=& C^2 \left[ \frac{\Gamma_1}{-2\alpha}\exp(-2\alpha t) +\frac{\Gamma_2}{-2\beta}\exp(-2\beta t) + \frac{\Gamma_1+\Gamma_2}{\alpha+\beta}\exp(-(\alpha+\beta)t)\right]^\infty_0 \nonumber\\
        &=&C^2 \left( \frac{\Gamma_1}{2\alpha}+\frac{\Gamma_2}{2\beta} - \frac{\Gamma_1+\Gamma_2}{\alpha+\beta} \right) \nonumber \\
        &=&\frac{1}{(\beta-\alpha)^2} \left( \frac{\Gamma_1}{2\alpha}+\frac{\Gamma_2}{2\beta} - \frac{\Gamma_1+\Gamma_2}{\alpha+\beta} \right) \nonumber\\
        &=& \frac{1}{(\beta-\alpha)}\left( \frac{\Gamma_1\beta - \Gamma_2 \alpha}{2\alpha\beta(\alpha+\beta)}\right) \nonumber
\end{eqnarray}

Therefore, $A=A_1-A_2$ is as follows.
\begin{eqnarray}
        A=\frac{1}{2\alpha\beta(\alpha+\beta)} \left( 1 - \frac{\Gamma_1\beta - \Gamma_2 \alpha}{\beta-\alpha} \right)
\end{eqnarray}

On the other hand, the RHS of eq. (10) denoted by $B$ is as follows.
\begin{eqnarray}
        B&=&k\int^{t'}_0 C(\exp(\alpha t)-\exp(\beta t)) dt \nonumber \\ 
        &=& C k \left[ -\frac{1}{\alpha}\exp(\alpha t) + \frac{1}{\beta} \exp(-\beta t) \right]_0^{t'} \nonumber\\
        &=& Ck \left( -\frac{1}{\alpha}(\Gamma_1-1) + \frac{1}{\beta}(\Gamma_2-1)\right) \nonumber \\
        &=& Ck \left( \frac{1-\Gamma_1}{\alpha}- \frac{1-\Gamma_2}{\beta}\right) \nonumber\\
        &=&\frac{k}{\beta-\alpha}\left( \frac{1-\Gamma_1}{\alpha}- \frac{1-\Gamma_2}{\beta}\right) \nonumber\\
        &=& \frac{k}{\alpha\beta}\left( 1-\frac{\Gamma_1\beta-\Gamma_2\alpha}{\beta-\alpha}\right). \nonumber
\end{eqnarray}

Therefore, $A=B$ implies that $\displaystyle k=\frac{1}{2(\alpha+\beta)}$.
\subsubsection*{Proof 4. Square pulse.}
Let $g(t)= a$ for $t<t_0$ and $g(t)= 0$ for $t>t_0$
The LHS of eq. \ref{eq:sup_toProof}, denoted by $A$, is
\begin{eqnarray}
        A &=&\int^\infty_0 g^2(t) dt - \int^\infty_0 g(t) g(t+t^{\prime}) dt \\
        &=& \int_{0}^{t_0} a^2 dt - \int_{0}^{t_0} a \cdot g(t+t^{\prime}) dt \\
        &=&a^2 t_0 - a\int_{0}^{t_0-t^{\prime}} a\; dt = a^2 t_0 -a^2 (t_0-t^{\prime}) \\ 
        &=& a^2 t^{\prime}
\end{eqnarray}
B, the RHS of eq. \ref{eq:sup_toProof} is
\begin{eqnarray}
        B&=&k\int_{0}^{\infty} g(t) dt= k\int_{0}^{t^{\prime}} a\; dt=k\; a\; t^{\prime} \\
        &=& k\; a\; t^{\prime} 
\end{eqnarray}

Therefore, $A=B$ implies that $\displaystyle k=a$ %ALGO MALO AQUI?

%---------------------------------------------------
\subsection*{$\boldsymbol{\Psi}$-patterns of uncorrelated pulse mixtures are linear combinations of the underlying pulse classes.}
Let us consider an extended definition of process $X$, which is composed of multiple classes of pulses $g_1(t), g_2(t),...g_n(t)$, with independent times of arrival (uncorrelated). For simplicity, the process can be stated by breaking it down into partial sums related to each pulse class ($N_1, N_2,..., N_n$ represent the number of pulses in each class).
\begin{equation}
X(t) = \sum^{N_1}_{j=1} g_1(t - t_j) + \sum^{N_2}_{j=1} g_2(t - t_j) + ... + \sum^{N_n}_{j=1} g_n(t - t_j)
\end{equation}

Given the independence of each subprocess (i.e. $cov(g_n, g_m) = 0$), the variance of $X$ is simply the sum of each subprocess's variances ($k$ is the index of each subprocess).
\begin{equation}
\sigma^2_X = \sum_k \operatorname{Var} \Big(\sum^{N_k}_{j=1} g_k(t - t_j)\Big)
\end{equation}
\noindent
and the variance of $\hat{X}$ is
\begin{equation}
\sigma_{\hat{X}}^2(t^{\prime}) = \sum_k \operatorname{Var}\Big(\sum^{N_k}_{j=1} g_k(t - t_j) - g_k(t - t_j + t^{\prime})\Big)
\end{equation}
\noindent
using the result shown in eq. \ref{eq:sup_varXtongo}
\begin{equation} \label{eq:sup_varXtongoLong}
\sigma_{\hat{X}}^2(t^{\prime}) = \frac{2}{f_s} \sum_k [\lambda_g]_k \Big( [E_g]_k - (g \star g)_k(t^{\prime}) \Big) = \sum_k [\sigma^2_{\hat{X}}]_k(t^{\prime})
\end{equation}

The terms $[\lambda_g E_g]_k$ and $[\lambda_g (g \star g)]_k(t^{\prime})$ define the energy contribution and the time-dependent component conferred by each pulse's class $g_k(t)$, respectively.

Finally, to recover the pulse classes, we employ the $\Psi$ operator (Eq. \ref{eq:PsiDef}), which requires us to differentiate eq. \ref{eq:sup_varXtongoLong}. Because differentiation is a linear operation, it can be distributed along the terms in the sum.
\begin{equation} 
\Psi_X(t^{\prime}) = \frac{1}{2} \sum_k \frac{d}{dt^{\prime}}[\sigma^2_{\hat{X}}]_k(t^{\prime}) = \sum_k [\Psi_X]_k(t^{\prime})
\end{equation}
\noindent
Thus, $\Psi_X(t^{\prime})$, which represents the global stochastic process, is a linear superposition of each subprocess's $[\Psi_X]_k(t^{\prime})$ (approximate profile of a class $k$ pulse), weighted by their energy contribution as stated in eq. \ref{eq:autoCov_sup}.

%-----------------------------------------------------------------------------------
\end{document}